\documentclass[letterpaper,11pt]{article}

\usepackage{graphicx}
\usepackage{fullpage}
\usepackage{xcolor}
\usepackage{mathtools}
\usepackage{slashed}
\usepackage{amsmath}
\usepackage{amssymb}
\graphicspath{{./figs/}}
\usepackage{dsfont}

\usepackage{xspace}

\newcommand {\e}{\epsilon}

\newcommand{\HS}{Hidden Sector\xspace}

\newcommand{\BR}{\text{Br}}

\definecolor{gray}{rgb}{0.8,0.8,0.8}

\title{Heavy Flavor \& Dark Sector}
\author{A.E.~Nelson and J.~Scholtz}

\begin{document}
\maketitle
\begin{abstract}
We consider some contributions to rare processes in B meson decays from a Dark Sector containing 2 light unstable scalars, with large couplings to each other and small mixings with Standard Model Higgs scalars. We show that existing constraints allow for an exotic contribution to high multiplicity final states with a branching fraction  as large as ${\cal O}(10^{-4})$, and that exotic particles could appear as narrow resonances  or long lived particles which are mainly found in high multiplicity final states from B decays.
\end{abstract}
\section{Introduction}

The last decade has seen an explosion of available measurements performed on the $B_d$ and $B_s$ meson systems. Their masses, mass differences, lifetimes, branching ratios of common and rare decays, asymmetries in their decays are all well measured. Unfortunately, there are very few deviations from the predictions put forth by the Standard Model  (SM), despite the effort poured into new sophisticated  methods   to interpret this data \cite{Amhis:2012bh,Laiho:2009eu,Bauer:2000yr}. With so few observed deviations, we are forced to wonder: Is this it? Is there any more physics we can extract out of B mesons? 

In other fields, such as Cosmology, we face puzzles of a different kind: A large body of evidence points towards the existence of Dark Matter (DM) as a significant ($\sim25\%$) component of our Universe. We know very little about DM: most of it is cold (from structure formation) and it interacts very weakly with itself (halo formation, bullet cluster) and with baryonic matter (direct detection, bullet cluster). The weakness of interaction between DM and SM particles justifies a separation of these two sectors. We will call the sector containing DM the Dark Sector (DS).

Although we know very little about dark matter, we know even less about the dark sector. The principle of Occam's Razor drives us towards the simplest theories of the DS with no additional particle content beyond what is necessary to explain the DM density of our Universe. However, this directly contradicts the nature of the Standard Model -- the degrees of freedom of the Standard Model far outnumber the degrees of freedom that participate in forming the 5\% of the Universe populated by baryonic matter. We  conclude that minimalism is not a valid principle for particle physics.

In this paper, we abandon minimalism and propose there are other fields and particles within the DS that do not contribute to the DM density of our Universe. There are a few reasons why only some DS particles might contribute significantly to DM density: some particles may freeze out at too low density, some might be too light to form a cold enough component of DM during the epoch of structure formation and some particles might be unstable on cosmological scales.

So far, the particles that form DM remain unobserved by direct detection experiments. Therefore, we wish to focus on the unstable particles that  do not contribute to the DM density. If their lack of stability comes from decay into   SM particles, we have a chance of observing their decay products in our detectors. Thus although the existence of DM motivates us to consider sectors which are weakly coupled to the SM, in this paper we do not discuss stable DM candidates at all.

Luckily, we have been given a physical system that is extremely sensitive to the existence of new decay channels. The B mesons, with their relatively long lifetimes and relatively low  mass, are ideally suited for probing  a GeV scale DS. Moreover, as already mentioned, these systems are very well explored by many dedicated experiments such as Belle, BaBar and LHCb as well as general purpose such as ATLAS and CMS. It would be a shame not to use this vast amount of experimental data to constrain the possible shape of the DS.

However, there are many different realizations of possible DS models and it is impossible to rule out all, or even a fraction of these models. In fact, a complete decoupling between the SM and the DS is a logical possibility that does not contradict any current experimental data, yet is impossible to rule out without a positive signal from the DS. Therefore, instead of focusing on constraining every corner of the DS model space, it would be far more fruitful to focus on describing possible signals that could arise as consequences of these models. This way we can alert the experimental community to measurements that may shed some light on the nature of these models. This approach is often called exploring the signal space, as opposed to exploring the model space.

Since we probe the B meson systems, it makes sense to use an effective field theory of the DS with a cut off above the B meson mass ($\sim 5$ GeV). In order to extract some interesting signals out of our DS, we have chosen to populate it with two scalars with internal couplings approaching a strongly coupled regime. If we wish, we can interpret these scalars as bound states of a strongly interacting theory or elementary scalars. In order to allow for some coupling between the two sectors, we include operators that contain both the Higgs fields and the DS scalars -- the so called Higgs Portal \cite{Schabinger:2005ei,Patt:2006fw,Strassler:2006im}.

We chose to include not just one SM Higgs field, but two Higgs Doublets \cite{Branco:2011iw}. This allows our models to be included not only within the Minimal Supersymmetric Standard Model (MSSM) frame work, but also allows us to use the decoupling limit which corresponds to the SM with one Higgs field. 

We discover that within our framework it is remarkably simple to significantly change the rate of rare decays of B mesons, in particular the decays into multi-particle final states. Depending on the parameters of our model these decays may appear prompt, or with displaced vertices. Finally, irrespective of including second Higgs doublets, light Higgs-like scalars preferentially couple to mesonic final states which motivates many new searches.

This paper is organized as follows: we will set up our model and establish our conventions and notation in section~\ref{sec:Setup}. We will present a UV completion of this model in section~\ref{sec:UV}. In section~\ref{sec:Ints} we will discuss the interaction between the SM and the DS. We will explore the experimental and theoretical constraints on our model in section~\ref{sec:constraints} and show the allowed branching fractions for high multiplicity decay modes of B mesons in section~\ref{sec:Bresults}. We finally conclude and suggest future directions in section~\ref{sec:conclusions}.

\section{Definitions, Notation and Setup}
\label{sec:Setup}

\subsection{The Model}

We will extend the SM in two ways. First, we use a   Higgs sector with two Higgs doublets. This is is a well known extension thanks to its presence in supersymmetric models. For the second extension, we take the simplest non-trivial low energy effective theory of the DS: two unstable scalars, both with masses on the order of GeV.  Although we call this sector the DS, we do not explicitly include the DM particle, as we are not assuming it is light enough to affect B decays.  We expect that all other dimension-full constants in this effective theory will be generated by the same processes and therefore will be roughly the same scale. The SM sector and the DS will be coupled through a Higgs Portal \cite{Schabinger:2005ei,Patt:2006fw,Strassler:2006im} -- a set of renormalizable operators that mix the 2HDM Higgs fields and the scalars in the DS. As a result, we split the Lagrangian into logically separate parts:
\begin{equation}
\mathcal{L} = \mathcal{L}_{\mathrm{SM}+\mathrm{2HDM}} + \mathcal{L}_{\mathrm{DS}} + \mathcal{L}_{\mathrm{Portal}},
\label{eq:}
\end{equation}
and discuss the individual parts in this section.

\subsection{Two Higgs Doublet  Extension of the Standard Model}

Two Higgs doublet extensions of the Standard Model are part of the standard lore of particle physics \cite{Branco:2011iw}. As opposed to the Standard Model (which we will occasionally call 1HDM), where only one Higgs field spontaneously breaks the Electro-Weak symmetry and gives mass to fermions, these extensions contain an additional Higgs doublet. In order to avoid large flavor-changing neutral currents, only one Higgs field is allowed to couple to up-type quarks, down-type quarks and leptons, respectively. We will use the type II model in which $H_u$ couples to the up-type quarks and $H_d$ couples to the down-type quarks and leptons. After Electro-Weak symmetry breaking (EWSB), this extension contains two massive neutral singlets $h_u$ and $h_d$. These mix and we will rotate the flavor basis $\{h_u,h_d\}$ into the mass eigenstate basis $\{h,H\}$:
\begin{equation}
	\left(\begin{matrix}h \\ H \end{matrix} \right) = \left( \begin{matrix} \cos \alpha & -\sin \alpha \\ \sin \alpha & \;\;\;\cos \alpha 				\end{matrix}\right) \left( \begin{matrix} h_u \\ h_d \end{matrix}\right)
\end{equation}
The ratio of the two vacuum expectation values of the two Higgs fields is called $\tan \beta = v_u/v_d$. The couplings of the light $h$ and the heavy $H$ to up-type and down-type fermions are then proportional to:
\begin{equation}
	 y_{u}^h = \frac{m_u}{v} \frac{\cos \alpha}{\sin \beta}	\hspace{0.25in} y_{d,l}^{h} = -\frac{m_{d,l}}{v} \frac{\sin \alpha}{\cos \beta} 	\hspace{0.25in} y_{u}^{H} = \frac{m_u}{v} \frac{\sin \alpha}{\sin \beta} \hspace{0.25in} y_{d,l}^{H} = \frac{m_{d,l}}{v} \frac{\cos \alpha}{\cos \beta} 
	\end{equation}
The 2 Higgs Doublet Model (2HDM) extension also contains a pseudoscalar neutral boson $A$ and a charged $H^\pm$, but they do not significantly contribute to our analysis since $H^\pm$ is charged and therefore does not mix with the DS and $A$ is typically too heavy under current experimental constraints.

\subsection{The Dark Sector}
As we state in the introduction, there is no reason why the DS should be simple. This view certainly complicates our ability to fully classify the effects of DS on measurable quantities. We take the view that although there is no reason for the DS to be simple, it is certainly preferable to start with a simple one. However, if too simple, the DS is unlikely to produce any novel  signature. In order to avoid both problems we take what we consider a minimal low energy effective theory of the DS which has distinctive consequences of multiple particle content. It contains two real scalars $n_1$ and $n_2$. We assume no symmetry properties for these scalars. This DS can be summarized by its Lagrangian\footnote{We assume the renormalized couplings are such that there is a stable vacuum at the origin of field space. This will constrain a combination of  the quadratic, quartic   and cubic terms.}:

\begin{equation}
\mathcal{L}_{\mathrm{DS}} = \frac{1}{2}\partial^\mu n_1 \partial_\mu n_1 + \frac{1}{2}m_1^2 n_1^2 + \frac{1}{2}\partial^\mu n_2 \partial_\mu n_2 + \frac{1}{2}m_1^2 n_2^2 + \frac{1}{3!}\sum_{ijk} \Lambda_{ijk} n_i n_j n_k + \frac{1}{4!}\sum_{ijkl} \lambda_{ijkl} n_i n_j n_k n_l 
\label{eq:DS}
\end{equation}   

In the next paragraph we will choose benchmark values of $m_1$, $m_2$ as well as $\Lambda_{ijk}$. We propose several mass study points for this DS as indicated in table~\ref{tab:SPs}. Study points SP1 and SP4 feature a particularly wide $n_1$. Currently, rather large values of $\e_1$ are allowed for $m_1 = 2\;\mathrm{GeV}$, which is why we choose three of the study points along this line (SP1,SP2,SP3). For completeness we also choose SP4 because it is a good representative for the low mass DS.

\begin{table}[ht]
	\begin{center}
		\begin{tabular}{l|c|c}
			Study Point & $m_1$ [GeV] & $m_2$ [GeV] \\ \hline \hline
			SP1 & $2.0$ & $0.85$ \\
			SP2 & $2.0$ & $0.5$ \\
			SP3 & $2.0$ & $0.3$ \\
			SP4 & $0.7$ & $0.3$
		\end{tabular}
	\end{center}
	\label{tab:SPs}
	\caption{List of Study Points.}
\end{table}

In order to avoid the existence of easily detected sharp resonances we require that the decay width for the process $n_1 \to n_2 n_2$ be as large as possible. We parametrize the dimensionful cubic in the following way:
\begin{equation}
\Lambda_{122} = \sqrt{16\pi\lambda_{122}} m_2
\end{equation}
The $ n_1 (n_2)^2$ operator is also responsible for mass correction to both $n_1$ and $n_2$, which is why we express it in terms of $m_2$. This way it is easier to track the contribution of $\Lambda_{122}$ to renormalization of $m_2$. With this parametrization, the width of $n_1$ takes a simple form:
\begin{equation}
\Gamma(n_1 \to n_2 n_2) = \lambda_{122} \frac{m_2^2}{m_1} \sqrt{1-4\frac{m_2^2}{m_1^2}}
\end{equation}
This is maximized for $m_1 = \sqrt{6}m_2$, leading to $\Gamma_1/m_1 \sim \lambda_{122}/10$. When $\lambda_{122}$ is large this theory becomes strongly coupled and our perturbative approach fails to make any sense. Also, for large enough $\lambda_{122}$ the cut-off needed to regulate the mass of $n_2$ becomes very low. We estimate that the boundary between the weakly coupled and the strongly coupled regimes sits around $\lambda_{122} \sim 1$ for $m_1 \sim m_2$, whereas the cut-off becomes too low ($\sim m_{B_s}$) at around $\lambda_{122} \sim 1/3$. Allowing a $1\%$ fine-tuning for $m_2^2$, $\lambda_{122}$ can be as large as $30$ -- far in the nonperturbative regime. Therefore as long as we stay within the perturbative regime, we do not have to be worried about fine-tuning between the cubic operators and the mass operator. For more details please read appendix~\ref{sec:widen1}.

\subsection{Higgs Portal}

As already advertised we will establish interactions with the Standard Model through the 2HDM generalized Higgs Portal. We will consider the set of all $2$ dimensional operators that cause mixing between 2HDM and DS scalars:
\begin{equation}
\begin{split}
\mathcal{L}_{\mathrm{Portal}} =&\; m_{1u}^2 h_u n_1 + m_{2u}^2 h_u n_2 + m_{1d}^2 h_d n_1 + m_{2d}^2 h_d n_2
\end{split}
\end{equation} 
In a general model we would have to find the eigenvectors of the full four dimensional ($\{h_u,h_d,n_1,n_2\}$) Hamiltonian. However, since we do not expect the cross-terms $m_{ix}^2$ to be very large, it is sufficient to define pairwise rotations by angles
\begin{equation}
\theta_{ix} = \frac{1}{2}\tan^{-1} \left( \frac{2 m_{ix}^2}{m_x^2-m_i^2}\right).
\end{equation}
These define the almost eigenstates $\tilde{n}_i$ and $\tilde{h}_x$:
\begin{equation}
\left(\begin{matrix} \tilde{h}_u \\ \tilde{n}_1 \end{matrix} \right)  = \left( \begin{matrix} \cos \theta_{u1} & \sin \theta_{u1} \\ -\sin \theta_{u1} & \;\;\;\cos \theta_{u1} \end{matrix}\right) \left( \begin{matrix} h_u \\ n_1\end{matrix} \right).
\end{equation}
We define $\theta_{2u}, \theta_{1d}$ and $\theta_{2d}$ similarly. The rotations defined by these angles do not commute, and so any successive application of these four rotations will not lead to mass eigenbasis of the model. However, as we will see in the subsequent sections, these angles are small and so all the terms arising from commutators are going to be suppressed and the states $\tilde{n}_i$ and $\tilde{h}_x$ are going to be for all practical purposes the eigenstates of the Hamiltonian. Ignoring the $h_u h_d$ mass mixing operator for now, we can use a single matrix to rotate into the mass eigenstate basis. To the first order in $\theta_{ix}$ this matrix takes a simple form:
\begin{equation}
\left(\begin{matrix}\tilde{h}_u \\ \tilde{h}_d \\ \tilde{n}_1 \\ \tilde{n}_2 \end{matrix} \right) = \left( \begin{matrix} 1 & 0 & \theta_{u1} & \theta_{u2} \\ 0 & 1 & \theta_{d1} & \theta_{d2} \\ -\theta_{u1} & -\theta_{d1} & 1 & 0 \\ -\theta_{u2} & -\theta_{d2}& 0 & 1 \end{matrix}\right)\left(\begin{matrix} h_u \\ h_d \\ n_1 \\ n_2 \end{matrix}\right)
\end{equation}
It is more convenient to express these angles by a different set of parameters:
\begin{equation}
	\begin{split}
	\theta_{u1} = \e_1 \cos \delta_1\\
	\theta_{d1} = \e_1 \sin \delta_1\\
	\theta_{u2} = \e_2 \cos \delta_2\\
	\theta_{d2} = \e_2 \sin \delta_2.
	\end{split}
\end{equation}
This way $\e_i$ stand for the amount of mixing between $n_i$ and the SM  Higgs fields, while $\tan \delta_i$ marks the ratio between $n_i$'s couplings with up-type and down-type fermions. In this treatment we only need to require that $\e_1,\e_2 \ll 1$ in order to ensure that all four mixing angles are small. Rotating into the mass eigenstate basis also introduces new mixed cubic and quartic operators between the two sectors. For example, we encounter a new operator that allows Higgs decay into a pair of DS scalars:
\begin{equation}
\mathcal{L}_{new} = \ldots + \frac{1}{2} \e_1 \cos \delta_1 \Lambda_{122} h_u n_2 n_2 + \ldots
\label{eq:cubics}
\end{equation} 
We will explore how this affects the range of allowed parameters in the later sections of this paper.

\section{ A UV Example with Naturally Light Hidden Scalars}
\label{sec:UV}
Although the model we have presented is mathematically consistent and renormalizable, it is interesting to consider whether there could be a natural origin for the small size of the scalar masses and the large size of their self couplings. 
We present an example in which they are composite particles, with naturally light masses. We take the two Higgs doublet model and populate the DS with a fermion $\psi$ that transforms under an $SU(N)$ with a confinement scale $\Lambda_D$. We add a heavy DS Higgs-like scalar $X$, with a vev $v_X$. The $X$ and Higgses mix and the UV Lagrangian for this DS takes the familiar form:
\begin{multline}
\mathcal{L} = \bar{\psi}\slashed{D}\psi + \lambda_\psi X \bar{\psi}\psi + \lambda_X \left(X^\dagger X- v_X^2\right)^2 +\\ +\lambda_{Xu}\left(X^\dagger X-v_X^2\right)\left(H_u^\dagger H_u - v_u^2\right)+\lambda_{Xd}\left(X^\dagger X-v_X^2\right)\left(H_d^\dagger H_d - v_d^2\right)
\end{multline} 
After symmetry breaking in the DS, we can integrate out the heavy $X$:
\begin{equation}
\mathcal{L} = \bar{\psi}\left(\slashed{D}+m_\psi\right)\psi + \cdots + \lambda_{Xu}\frac{\lambda_\psi v_X \bar{\psi}\psi }{M_X^2}\left(H_u^\dagger H_u - v_u^2\right) + (u \longleftrightarrow d)
\end{equation}
Below $\Lambda_D$, the $\psi$ are confined into mesons: $\bar{\psi}\psi \to f_{D}^2 n_i$. Thus we get an effective field theory for a bound state of $\bar{\psi}\psi$ coupled to our Higgses:
\begin{equation}
\mathcal{L} = (\partial n_i)^2 + m_i^2 n_i^2 + \left(\frac{\lambda_{Xu}\lambda_{\psi} v_X v_u f_{D}^2}{M_X^2}\right) h_u n_i + (u \longleftrightarrow d),
\end{equation} 
which corresponds to a misalignment between the flavor and mass basis of the order:
\begin{equation}
\theta_i^u \sim \frac{\lambda_{Xu}\lambda_{\psi} v_X v_u f_{D}^2}{M_X^2 m_h^2} \leq \frac{\lambda_{Xu} m_{i} v_u f_{D}^2}{M_X^2 m_h^2} \sim \frac{\lambda_{Xu} m_{i}^3 v_u }{M_X^2 m_h^2} \frac{f_{D}^2}{m_{i}^2} \sim \frac{\lambda_{Xu} m_{i}^3 \cos \beta }{M_X^2 m_h} \frac{f_{D}^2}{m_{i}^2}
\end{equation}
Suppose that $X$ is not much heavier than $m_h$, then we expect:
\begin{equation}
\theta_i^u \sim 10^{-5} \cos \beta \left(\frac{ \lambda_{Xu}}{0.1}\right) \left(\frac{f_{D}}{m_{i}}\right)^2
\end{equation}
However, if the $SU(N)$ coupling remains strong between $M_X$ and $m_i$, the operator $\bar{\psi}\psi$ might have a large anomalous dimension $\gamma \sim \mathcal{O}(1)$ near such an infrared conformal fixed point. This means that the operator
\begin{equation}
	\left(\frac{\lambda_{Xu}\lambda_{\psi} v_X v_u f_{D}^2}{M_X^2}\right) h_u n_i
\end{equation}
would be scaled by a factor:
\begin{equation}
	\left(\frac{M_X}{m_i}\right)^\gamma.
\end{equation}
This would allow a much larger $\theta_i^u  \sim 10^{-1}$. It is possible to double the Dark Higgs sector in order to allow for different couplings between the DS bound states and Standard Model up and down Higgses. 

Note that in this model there will be other states besides our minimal pair of scalars. As long as it contains a scalar $n_1$ that can decay into 2 mesons $n_2$, which in turn are unable to decay into any hidden states, the signatures we discuss could be present. As $\psi$ number is conserved,   there will be a new stable dark ``baryon", which is a bound state of $N$ $\psi$ particles, and is a dark  matter candidate. As this baryon is heaver than the scalars by a factor of ${\cal O}(N)$, we assume it does not appear in
 B meson decays.

\section{Interactions between the Dark Sector and the Standard Model}
\label{sec:Ints}
	We would like to observe measurable effects of our model in decays of B mesons. Therefore, we need to make sure B mesons can decay into the DS. Moreover, unless we want to look for events with just missing energy we also need make sure that the DS particles decay back into Standard Model particles. In the next two sections we show how this can be done.

\subsection{B decays through the Higgs penguin}
	We are interested in B meson decays into the DS. This happens through the Higgs penguin operator $\bar{s}bh$ and the Higgs Portal. The Standard Model Higgs penguin has a relatively simple form compared to its 2HDM cousin. In the 2HDM extension the total size of the matrix elements as well as the ratio between the $\bar{s}bh$ and $\bar{s}bH$ couplings are functions of the form of the 2HDM extension as well as $\tan \beta$ and $\alpha$. We will parametrize this model dependence by two parameters, $\xi$ and $\gamma$, that modify the SM operator:
\begin{equation}
		\begin{split}
		\mathcal{L}_{bs} =& \frac{3\sqrt{2}G_F m_t^2 V_{ts}^*V_{tb} \xi(\tan \beta, m_t,m_W,\ldots) }{16\pi^2 v } \left( h \cos \gamma + H \sin \gamma \right)[\bar{s}_L b_R]\\
		 =& \xi \lambda_p	\left( h \cos \gamma + H \sin \gamma \right)[\bar{s}_L b_R],
		 \end{split}
	\end{equation}
where we have defined:
\begin{align}
 \lambda_q &= \frac{3\sqrt{2}G_F m_t^2 V_{tq}^*V_{tb} m_b}{16\pi^2 v }\\
 |\lambda_s| &= 9.47\times 10^{-6}\\
 |\lambda_d| &= 1.85\times 10^{-6}
\end{align}
Notice that these parameters are degenerate with other parameters in our model. For example, take the coupling $\bar{s}b n_i$:
\begin{equation}
\begin{split}
\xi \lambda_p	\left( h \cos \gamma + H \sin \gamma \right)[\bar{s}_L b_R] = \xi \lambda_p	\left( h_u \cos (\gamma-\alpha) + h_d \sin(\gamma-\alpha) \right)[\bar{s}_L b_R] = \\
 = \lambda_p (\xi  \e_i)	 \cos (\gamma-\alpha-\delta_i)  n_i[\bar{s}_L b_R]+\ldots,
\end{split}
\end{equation}
and so until we have detailed knowledge of the 2HDM Higgs sector\footnote{A Supersymmetric 2HDM will give different penguin strength compared to a simpler 2HDM extension} we will be content with expressing all predictions in terms of $\xi \e_i$ and $\delta_i$.

The LHC has discovered a 126 GeV Higgs particle which is  Standard Model-like \cite{Chen:2013rba}. These studies strongly prefer $\sin(\alpha-\beta)=1$ and allow a somewhat large range for $\tan \beta$ including $\tan\beta = 1$. This would foreshadow Standard Model-like penguin diagrams with $\xi \sim 1$ and $\gamma \sim 0$. We will, for study purposes, use these values. However, due to the above mentioned degeneracy even if these are not correct assumptions our study can be easily recast into different 2HDM scenarios.

The nature of the link between the Standard Model B mesons and the DS scalars implies correlations between different decay channels, which should be exploited when identifying this particular DS. For example, an excess of events in $B_d \to K^0 \mu\mu$ should be accompanied by a similar excess in $B^\pm \to K^\pm \mu\mu$, $B_s \to \phi \mu\mu$ as well as a smaller excess (by a factor of $|V_{td}/V_{ts}|^2$) in $B_s \to K \mu\mu$. Similarly, an excess in $B_s \to 4\pi$ should come with a similar excess in $B_d \to K+4\pi$ and $B_s \to \phi+4\pi$.

\subsection{Decays of $n_1$ and $n_2$}

We have already ensured that $n_1$ decays very quickly into two $n_2$s by setting $\lambda_{122}$ as large as possible. However kinematic constraints only allow $n_2$ to decay into Standard Model particles. Its couplings through the Higgs Portal allow decays into pairs of leptons, mesons and photons. Given the nature of its couplings, the branching fractions into these modes are identical to those of a light Higgs boson and are dependent on the mass of $n_2$ as well as $\delta_2$.

Ordinarily, for $m_2 < 2m_K$, we could be content with the chiral perturbation theory ($\chi$PT) prediction featured in appendix~\ref{sec:branch}. However, Donoghue \emph{et al.} have shown in \cite{Donoghue} that higher order contributions generate a non-zero $\Delta_\pi = \langle \pi\pi | \bar{s}s | 0\rangle$ matrix element (which violates the OZI rule). The coefficient of this operator, $m_s$, is large enough to make its contribution towards $n_2 \to \pi\pi$ significant. One can think about this contribution as creating a virtual pair of kaons that rescatter into a pair of pions.

\begin{figure}%
\centering
\includegraphics[width=0.5\textwidth]{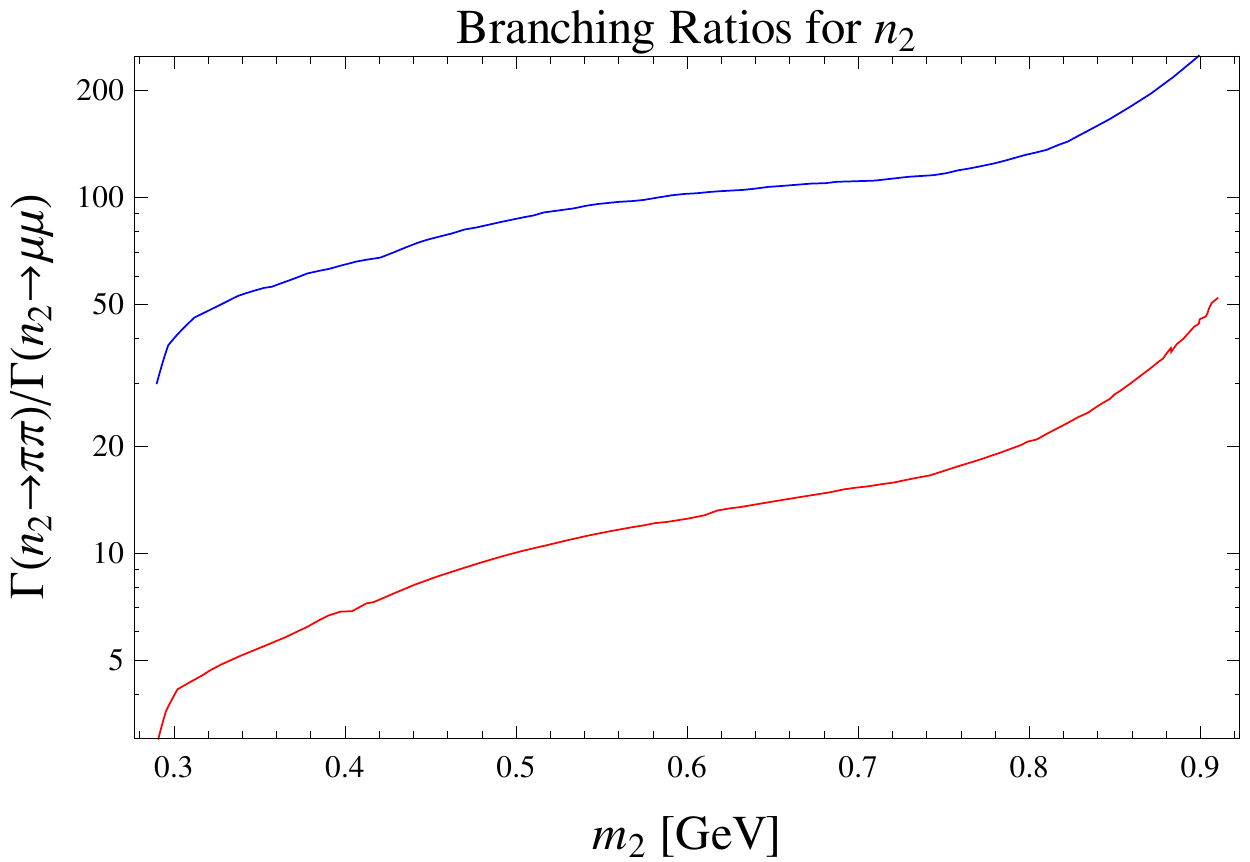}
\caption{Branching ratio $\BR(n_2\to\pi\pi)/\BR(n_2\to\mu\mu)$ as a function of $m_2$. The lower, red curve corresponds to $\delta_2 = \pi/4$, the higher, blue curve represents the choice $\delta_2=pi/16$.}%
\label{fig:n2br}%
\end{figure}

We will use data from \cite{Donoghue} to form  a more complete picture of decays of $n_2$. However,   close to the $m_2 = 2m_K$ threshold, where the ratio $\BR(\pi\pi)/\BR(\mu\mu)$ is significantly enhanced, the  approximations used may not be very reliable, and the predictions in this mass region should be taken with a grain of salt. The authors of \cite{Donoghue} separate the transition operator into three parts:
\begin{equation}
\langle \pi\pi | \mathcal{O} | 0 \rangle = \langle \pi\pi | \theta_\mu^\mu | 0 \rangle + \langle \pi\pi | m_s \bar{s}s | 0 \rangle + \langle \pi\pi | m_u \bar{u}u+m_d\bar{d}d | 0 \rangle = \theta_\pi + \Delta_\pi + \Gamma_\pi
\end{equation}
The contribution from $\Gamma_\pi$ (due to smallness of $m_u$ and $m_d$) is negligible and we will omit this operator in our analysis. In our model the couplings to up and down fermions are modified to:
\begin{equation}
m_u\bar{u}u \rightarrow \e_2\frac{\cos \delta_2}{\sin \beta} m_u \bar{u}u \hspace{0.5in} m_{d}\bar{d}d \rightarrow \e_2\frac{\sin \delta_2}{\cos \beta} m_d \bar{d}d \hspace{0.5in} m_{l}l^+l^- \rightarrow \e_2\frac{\sin \delta_2}{\cos \beta} m_l l^+l^-,
\end{equation}
which means that the relative branching fraction between pairs of pions and muons depends on $\beta$ and $\delta_2$:
\begin{equation}
\frac{\Gamma(n_2\to\pi\pi)}{\Gamma(n_2\to\mu\mu)}= \frac{\left|\left(\frac{2 \cot \delta_2 \cot \beta + 1}{3}\right)\mathrm{BF}(SM,\Delta_\pi = 0)^{\frac{1}{2}} + \left(\frac{25-4\cot \delta_2 \cot \beta}{21}\right)\mathrm{BF}(SM,\theta_\pi = 0)^{\frac{1}{2}} \right|^2}{\mathrm{BF}(n_2\to\mu\mu)_{SM}},
\end{equation}
where $\mathrm{BF}(SM,\mathcal{X}=0)$ is the branching fraction for a Standard Model Higgs with the operator $\mathcal{X}$ turned off, while $\mathrm{BF}(n_2\to\mu\mu)_{SM}$ would be the branching fraction of $n_2$ in a model with single Higgs boson. Since the phases of $\theta_\pi$ and $\Delta_\pi$ are identical, we can extract exact contribution of each operator \cite{Donoghue}. Figure~\ref{fig:n2br} shows our results for the branching ratio $\Gamma(n_2\to\pi\pi)/\Gamma(n_2\to\mu\mu)$ for a range of $m_2$.

\section{Constraining the Model}
\label{sec:constraints}
In order to make our task manageable we will limit the range of some parameters (such as $m_1$ and $m_2$) as well as only use a set of discrete values for other parameters ($\delta_1$ and $\delta_2$). Using chosen values we will then derive constraints on $\e_1$ and $\e_2$. With a complete set of parameters we will then make predictions for multi-particle final states that have not been yet measured and (happily) point out that the allowed rates are large and (hopefully) observable.

How do we extract $\e_1$ and $\e_2$? First, in agreement with our initial desire to work with an almost strongly coupled DS we set all the DS scalar couplings:
\begin{align}
\Lambda_{111} = \sqrt{16\pi\lambda_{111}}m_1,\;\;\;\lambda_{111}=1\\
\Lambda_{112} = \sqrt{16\pi\lambda_{112}}m_2,\;\;\;\lambda_{112}=1\\
\Lambda_{122} = \sqrt{16\pi\lambda_{122}}m_2,\;\;\;\lambda_{122}=1
\end{align}
Since operators $\Lambda_{112}n_1^2 n_2$ and $\Lambda_{122}n_1 n_2^2$ contribute to renormalization of $m_2$ we make them proportional to $m_2$. On the other hand $\Lambda_{111}$ is proportional to $m_1$ since it does not renormalize $m_2$ at one-loop level.

The processes $B_q \to M \mu\mu$ and $B_q \to M \pi\pi$ are dominated by the narrow $n_2$ resonance and their rates are virtually independent of any of the properties of $n_1$. Therefore, we use these processes to constrain $\e_2(m_2,\delta_2)$. The allowed $\e_2(m_2,\delta_2)$ is low enough that New Physics contribution to processes such as $B_q \to \mu\mu$, $B_q \to \pi\pi$ as well as $B_q \to 4\mu$ or $B_q \to 4\pi$ is dominated by $n_1$ in the $s$-channel. This means we can use the two and four body decays of $B_q$ to constrain $\e_1(\delta_1,\delta_2,m_1,m_2)$ for given $\delta_1$, $\delta_2$, $m_1$ and $m_2$.

We still must  specify $\delta_1$ and $\delta_2$. We choose $\delta_1 = \pi/4$. Although we could choose a different value, present constraints on this DS do not force us to go beyond the simplest case.

The parameter $\delta_2$ determines whether the final states of DS decays are hadronic or leptonic. We choose two different scenarios: the 1HDM equivalent $\delta_2 = \pi/4$ and the somewhat leptophobic $\delta_2 = \pi/16$. We believe that possibly the best motivation for the somewhat leptophobic scenario is that it represents a logical possibility that provides a motivation for exploring a large swath of experimental scenarios such as high multiplicity hadronic final states. However, note that $\cot \pi/16 \sim 5$, therefore this is not a particularly fine-tuned scenario. With every other parameter in place we are ready to constrain $\e_1$ and $\e_2$. 

\subsection{Constraining with $\Upsilon$ decays}
The  branching fraction for a heavy vector   state $\Upsilon$ to decay into a photon and a very light higgs particle with mass $m_i$ was estimated in ref. \cite{Wilczek:1977zn} to be 
\begin{equation}
 \frac{G_F m_b^2}{\sqrt{2} \pi\alpha}\left(1-\frac{m_i^2}{m_\Upsilon^2}\right) {\rm Br}\left( \Upsilon\rightarrow\mu\mu\right).
\end{equation}
Since the light hidden scalars mix with $H_d$,
the $\Upsilon$ could decay into a photon and either $n_1$ or $n_2$, with a branching fraction suppressed by an additional factor of of the mixing angle $\theta_{di}$ squared, and enhanced by $\tan^2\beta$. Since for light scalars this constraint on the parameter space is less stringent than the constraints from B mesons we will not consider it further.
\subsection{Constraining $\e_2$ with Three Body Final States}

Decays of B mesons into a meson and $n_i$ result in final states such as $K\mu\mu$, $K^*\mu\mu$, $\phi \pi\pi$. The $s$-channel contribution from the broad $n_1$ is negligible compared to the much narrower on-shell $n_2$ as long as $\e_2 > 10^{-4}\e_1$, which we will find to be true. Therefore, these decay channels only depend on $\e_2$, $\delta_2$ and $m_2$. Since many of these final states are well constrained by experimental measurements and some are accessible to theoretical predictions with varying range of accuracy and reliability we can use these measurements and predictions to put significant constraints on $\e_2$. Table~\ref{tab:BtoMn2} lists the decay channels we use to constrain our model as well as the HFAG combinations \cite{Amhis:2012bh}, the Standard Model predictions and the allowed $2\sigma$ deviation for each channel. Similar results in agreement with ours can be found in \cite{Batell:2009jf}.
\begin{table}[ht]
\centering
\begin{tabular}{lccc}
\hline
Process & HFAG combination \cite{Amhis:2012bh} & SM prediction & Allowed $2\sigma$ Excess \\\hline
$B_d \to K \mu\mu$, & $0.32^{+0.21}_{-0.20} \times 10^{-7}$ & $(0.67\pm 0.28)\times 10^{-7}$, \cite{Bouchard:2013mia}& $0.35\times 10^{-7}$ \\
\hspace{0.25in}$q^2 < 2\mathrm{GeV}^2$& & & \\
$B_d \to K\pi\pi$ (NR)& $(14.7\pm 2.0)\times 10^{-6}$ & Unreliable &  $18.7\times 10^{-6}$ \\
$B_d \to K^*\mu\mu$,& $(1.46\pm 0.5)\times 10^{-7}$ & $(2.0\pm0.25)\times 10^{-7}$, \cite{Beneke:2001at} & $0.57\times 10^{-6}$\\
\hspace{0.25in}$q^2 < 2\mathrm{GeV}^2$& & & \\
$B_d \to K^*\pi\pi$ & $(55\pm5)\times 10^{-6}$ & Unreliable & $65\times 10^{-6}$ \\\hline
$B_u \to K^+ \mu\mu$,& $(0.53\pm0.04) \times 10^{-7}$ &  $(0.67\pm 0.28)\times 10^{-7}$, \cite{Bouchard:2013mia}& $0.42\times 10^{-7}$\\
\hspace{0.25in}$q^2 < 2\mathrm{GeV}^2$& & & \\
$B_u \to K^+\pi\pi (NR)$ & $(16.3 \pm 2.0) \times 10^{-6}$ & Unreliable & $20.3\times 10^{-6}$\\
$B_u \to K^{+*}\mu\mu$ & $(1.41\pm0.5)\times 10^{-7}$ & $(2.0\pm0.25)\times 10^{-7}$, \cite{Beneke:2001at} & $0.52\times 10^{-6}$  \\
\hspace{0.25in}$q^2 < 2\mathrm{GeV}^2$& & & \\
$B_u \to K^{+*}\pi\pi$ & $(75.3\pm10.1)\times 10^{-6}$ & Unreliable & $95.5\times 10^{-6}$\\\hline
$B_s \to \phi \mu\mu$ & $(0.91\pm0.24)\times10^{-6}$ & $1.23\times 10^{-6}$ \cite{Geng:2003su} & $0.16\times10^{-6}$\\
$B_s \to K \pi\pi$ & $(11.9\pm3.7)\times10^{-6}$ & Unreliable & $19.3\times 10^{-6}$\\\hline
\end{tabular} 
\caption{Some three body decay channels of B mesons we use to constrain the parameters of our model. NR stands for non-resonant and $q^2 < 2\text{ GeV}^2$ implies a constraint in a particular bin of the differential cross-section.}
\label{tab:BtoMn2}
\end{table}
Every $B_d$ channel has an equivalent $B_u$ channel. The currents responsible for these transitions are identical (if we treat the $u$ and $d$ quarks as spectators there is no difference at all) and so up to minor electromagnetic corrections these modes are nearly identical. The experimental constraints are also very similar and so we list the charged B meson modes for completeness rather than for additional information. Notice that for the same reason the lattice predictions are identical for the neutral and charged modes.
\begin{figure}[ht]%
\centering
\includegraphics[width=0.45\textwidth]{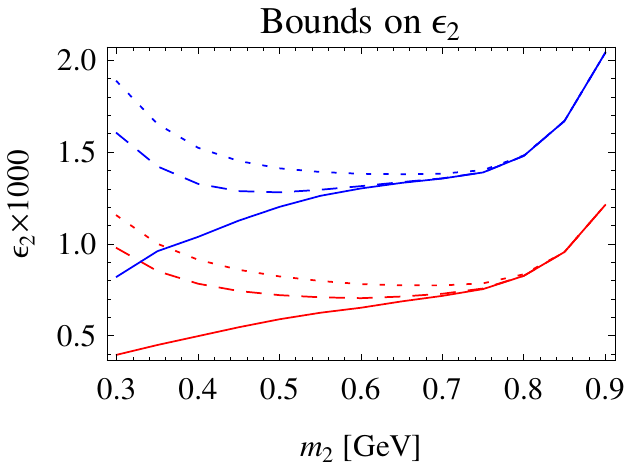}\hspace{0.2in}\includegraphics[width=0.45\textwidth]{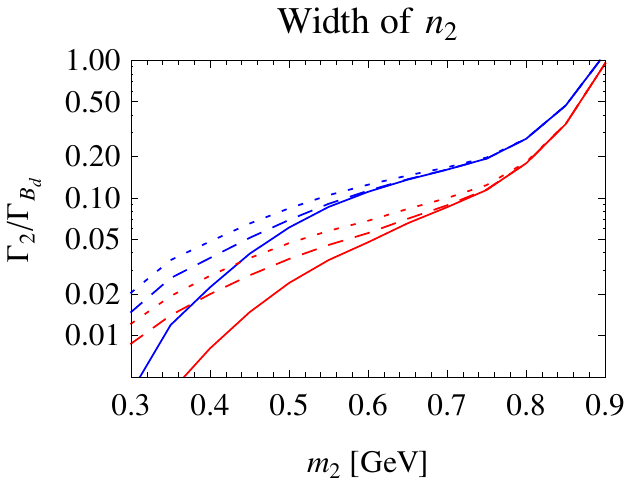}
\caption{On the left, the bounds on $\e_2$ for $\delta_2=\pi/4$ are shown in red and the bounds for $\delta_2=\pi/16$ are shown in blue. The solid lines represent results of a naive analysis that assumes a full sensitivity independent of the lifetime of $n_2$. The dotted and dashed lines represent the bounds on $\e_2$ if decays that happen within $5\;\mathrm{cm}$ or $10\;\mathrm{cm}$ of the primary interaction point were recorded. The right figure shows the width of $n_2$ in terms of width of $B_d$. The dashed lines show the actual values including the correction from lowered detector acceptance due to long lifetimes of $n_2$.}%
\label{fig:e2plot}%
\end{figure}
The widths for $B_q \to P n_2$ and $B_q \to V n_2$ are expressed in terms of Form Factors adopted from \cite{Ball:2004ye,Ball:2004rg}:
\begin{multline}
\Gamma(B_q \to P n_2) = \frac{\lambda_q^2 \e_2^2 \cos^2(\gamma-\alpha-\delta_2)}{64\pi m_{B_q}^3 m_b^2}\left(m_{B_q}^2-m_P^2\right)^2 \left|f_0^{q\to P}(m_2^2)\right|^2 \times\\
\times \left[\left(m_{B_q}^2-(m_2-m_P)^2\right)\left(m_{B_q}^2-(m_2+m_P)^2\right)\right]^{1/2}
\label{eq:Bto3a}
\end{multline}
\begin{equation}
\Gamma(B_q \to V n_2) = \frac{\lambda_q^2 \e_2^2 \cos^2(\gamma-\alpha-\delta_2)}{16\pi m_{B_q}^3 m_b^2} \left|A_0^{q\to V}(m_2^2)\right|^2 \left[\left(m_{B_q}^2-(m_2-m_V)^2\right)\left(m_{B_q}^2-(m_2+m_V)^2\right)\right]^{3/2}
\label{eq:Bto3b}
\end{equation}
Since $n_2$ is very narrow, there is virtually no interference between the SM processes and New Physics, which justifies incoherently adding results of equations~\ref{eq:Bto3a} and~\ref{eq:Bto3b} to the Standard Model contribution. The spectrum of the invariant mass of the two muons would show a narrow peak centered around $m_2$.  It may seem dangerous to place a very narrow line in a well measured process. However, although the differential width $d\Gamma/dq^2$ is a well measured quantity, a search for narrow lines has been only done for $m_2< 0.3\;\mathrm{GeV}$ \cite{Hyun:2010an}. Above this mass the results are quoted in somewhat coarser bins. Since our study points satisfy $m_2^2 < 2\;\mathrm{GeV}^2$, when the differential width measurements are available, we use the binned ($0<q^2< 2\;\mathrm{GeV}^2$) measurement to obtain stronger constraints on $\e_2$. 

When $\e_2$ is low enough, $n_2$ becomes long-lived on detector scales. As a result, the detector acceptance suffers and the bounds on $\e_2$ weaken. In order to model this effect when considering the bounds on $\e_2$ we only consider the portion of $n_2$s that decay within $5\;\mathrm{cm}$ or within $10\;\mathrm{cm}$ from the primary interaction point. We summarize these bounds on $\e_2$ in figure~\ref{fig:e2plot}.
 
\subsection{Constraining $\e_1$ with $B_q \to \mu\mu$ and $B_q \to \pi\pi$}

With recent experimental determination of the branching fraction $B_s \to \mu\mu$ and ever increasing constraints on $B_d \to \mu\mu$, these two channels could provide a constraint on our model. In $B_q \to n_i \to \mu\mu$ the momentum flowing through $n_i$ is fixed to $q^2 = m_B^2$. Unless $m_1$ or $m_2$ are close to the mass of the B meson, this processes is not enhanced by any resonances as it was in $B_q \to M n_i$. Given that the $n_1$ and $n_2$ propagators are both of nearly equal size, the relative strength of these two $s$-channel processes is set by the ratio $\e_1/\e_2$. However, bounds from $B_q \to M n_2$ force $\e_2$ so low that $n_2$ has no measurable effect on this branching fraction. Notice that the contribution from the neutral Higgs particle with mass $m_h$ is suppressed by $(m_B/\e_1m_h)^4 \sim (25 \e_1)^{-4}$. Therefore we cannot constrain $\e_1$ much below $0.04$ using this decay. As a result the expression for this partial width is relatively simple (as long as $\e_1\gtrsim 0.04$):
\begin{equation}
\Gamma(B_q \to n_1^* \to \mu\mu) = \frac{1}{8\pi}\frac{m_{B_q}^5 m_{\mu}^2 f_{B_q}^2 }{v^2 m_b^2 \cos^2 \beta}\frac{\lambda_q^2 \e_1^4 \cos^2(\gamma-\alpha-\delta_1)\sin^2(\delta_1)}{(m_{B_q}^2-m_1^2)^2 + m_1^2 \Gamma_1^2(m_{B_q}^2)},
\end{equation}
where it is important to evaluate the width of $n_1$ at $q^2 = m_B^2$. We show the experimental values and the Standard Model predictions for branching fractions as well as the allowed $2\sigma$ deviations for the decay modes of interest in table~\ref{tab:Bto2mu}.
\begin{table}[ht]
\centering
\begin{tabular}{lccc}
\hline
Process & HFAG combination \cite{Amhis:2012bh} & SM prediction & Allowed $2\sigma$ Excess \\\hline
$B_s \to \mu\mu$ & $(3.2\pm 1)\times 10^{-9}$ & $(3.23\pm 0.27)\times 10^{-9}$, \cite{Buras:2012ru} & $2 \times 10^{-9}$ \\
$B_s \to \pi\pi$ & $(0.73\pm0.14)\times 10^{-6}$ & $(0.57^{+0.26}_{-0.23})\times 10^{-6}$, \cite{Ali:2007ff} & $0.76\times 10^{-6}$\\ \hline
$B_d \to \mu\mu$ & $< 8\times 10^{-10}$ & $(1.07\pm0.1)\times 10^{-10}$, \cite{Buras:2012ru} & $7\times 10^{-10}$\\
$B_d \to \pi\pi$ & $7.01\pm 0.29$ & pQCD: $(6.7\pm3.8)\times 10^{-6}$, \cite{Ali:2007ff} & $7.9\times 10^{-6}$\\
 &  & SCET: $(6.2\pm4)\times 10^{-6}$, \cite{Williamson:2006hb} & $9.1\times 10^{-6}$\\ \hline
\end{tabular} 
\caption{Two particle decay channels of B mesons we use to constrain $\epsilon_1$.}
\label{tab:Bto2mu}
\end{table}
The constraints from these processes are in general not strong enough to constrain $\e_1$. This is because these processes do not create any on-shell DS states and are therefore suppressed by the additional factors of $(\e_1 m_\mu/v)^2 \sim 10^{-9}$. In general we will obtain much higher rates (at the possible cost of displaced vertices) by creating on-shell DS states that decay into SM particles later. The most constraining modes are presented in the next section.

\subsection{Constraining $\e_1$ with $B_q \to 4\mu$ and $B_q \to 4\pi$}

As we have mentioned, $B_q \to n_i \to \mu\mu$ do not constrain $\e_1$ all that well in comparison with other decay modes such as $B_q\to 4\mu$. Similar to the three particle final state, the dominant contribution to $B_q \to 4\mu$ comes from $B_q \to 2n_2 \to 4\mu$ with both $n_2$s on-shell. The width for this processes is not complicated:
\begin{multline}
\Gamma(B_q \to 2n_2) = \frac{1}{32\pi}\frac{\lambda_q^2 f_{B_q}^2 m_{B_q}^3 }{m_b^2}\left|\frac{4m_2\sqrt{\pi\lambda_{122}}\e_1 \cos(\gamma-\alpha-\delta_1)}{m_{B_q}^2-m_1^2+ i m_1 \Gamma_1(m_B^2)}+\frac{4m_2\sqrt{\pi\lambda_{222}}\e_2 \cos(\gamma-\alpha-\delta_2)}{m_{B_q}^2-m_2^2+ i m_2 \Gamma_2(m_B^2)}\right.+\\+\left.\frac{\Lambda_{h22}\cos(\gamma)}{m_{B_q}^2-m_h^2}+\frac{\Lambda_{H22}\sin(\gamma)}{m_{B_q}^2-m_H^2}\right|^2
\end{multline}
Notice that we have set $\lambda_{122} = \lambda_{222}$ and the $n_1$ and $n_2$ propagators are dominated by $m_B^2$ and so their relative contribution is again determined by the ratio $\e_1/\e_2$. We only need to keep the contribution from $n_1$ unless $\e_1 \sim \e_2$. The Higgs contribution is suppressed by
\begin{equation}
f_h = \frac{\Lambda_{h22} m_B^2}{\e_i\Lambda_{i22} m_h^2} \sim \frac{\Lambda_{h22}}{\e_i \Lambda_{i22}} \frac{m_B^2}{m_h^2}.
\end{equation}
We expect $\Lambda_{h22} \sim \e_i\Lambda_{i22}$ and so the $s$-channel Higgs contribution is suppressed by $(m_B/m_h)^2$ and is negligible. This is true every time $n_1$ carries momentum much smaller than the mass of Higgs and we will ignore the low momentum Higgs boson contribution in the future. Thus, we only need to consider a simple partial width:
\begin{equation}
\Gamma(B_q \to 2n_2) = \frac{\lambda_q^2 f_{B_q}^2 m_{B_q}^3 m_2^2}{m_b^2}\frac{\lambda_{122}\e_1^2 \cos^2(\gamma-\alpha-\delta_1)}{(m_{B_q}^2-m_1^2)^2+ m_1^2\Gamma_1(m_B^2)^2}
\end{equation}
The properties of the three experimentally measured decay channels that fall into this category are summarized in table~\ref{tab:Bto2n2}.
\begin{table}[ht]
\centering
\begin{tabular}{lccc}
\hline
Process & HFAG combination \cite{Amhis:2012bh} & SM prediction & Allowed $2\sigma$ Excess \\\hline
$B_s \to 4\mu$ & $< 1.2\times 10^{-8}$ & NR: $<10^{-10}$, \cite{Melikhov:2004mk} & $< 1.2\times 10^{-8}$ \\\hline
$B_d \to 4\mu$ & $<5.1\times 10^{-9}$ & Negligible & $<5.1\times 10^{-9}$ \\
$B_d \to 4\pi$ & $<19.3 \times 10^{-6}$ & Unreliable & $<19.3 \times 10^{-6}$\\\hline
\end{tabular} 
\caption{Four particle decay channels of B mesons we use to constrain $\e_1$.}
\label{tab:Bto2n2}
\end{table}
\begin{figure}[ht]%
\centering
\includegraphics[width=0.4\columnwidth]{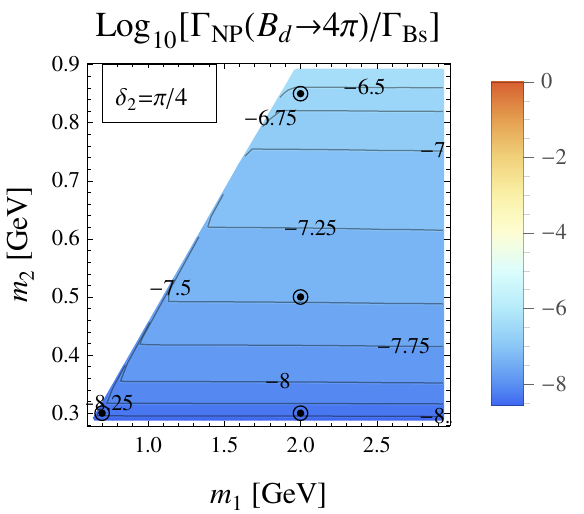}\hspace{0.05in}\includegraphics[width=0.4\columnwidth]{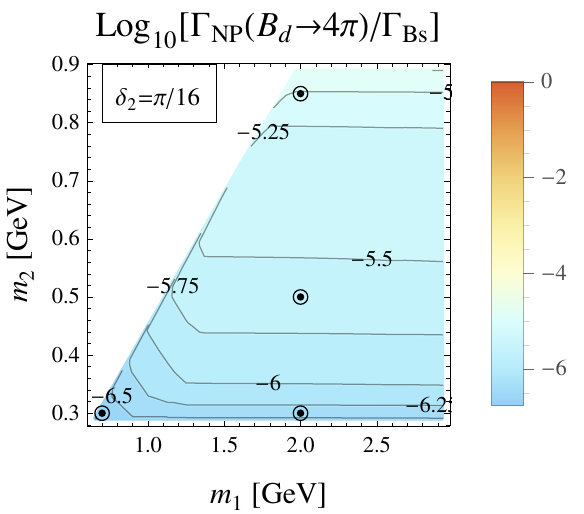}\\\includegraphics[width=0.4\columnwidth]{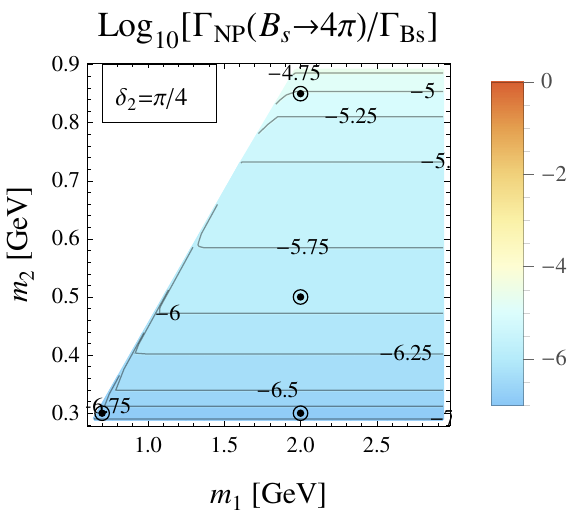}\hspace{0.05in}\includegraphics[width=0.4\columnwidth]{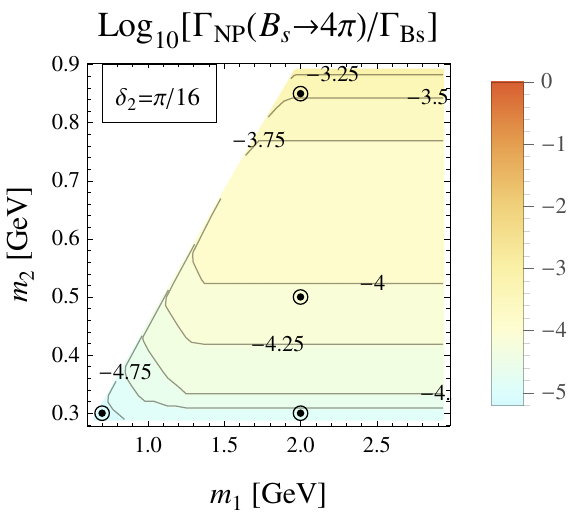}
\caption{Plots of $B_d \to 4\pi$ and $B_s \to 4\pi$. $B_d \to 4\pi$ comes close to saturating the experimental bound for $m_2 \sim 0.9\;\mathrm{GeV}$. However, exploring the branching fraction for $B_s \to 4\pi$ would significantly constrain the model.}
\end{figure}
The most stringent test (not surprisingly) comes from $B_s \to 4\mu$ since the other two channels are suppressed by $|V_{td}/V_{ts}|^2$. Measuring $B_s \to 4\pi$ would provide a great constraint on $\e_1$ for $\delta_2 = \pi/16$, however, an experimental measurement of this mode is currently unavailable.
 
\begin{figure}[ht]
\centering
\includegraphics[width=0.4\textwidth]{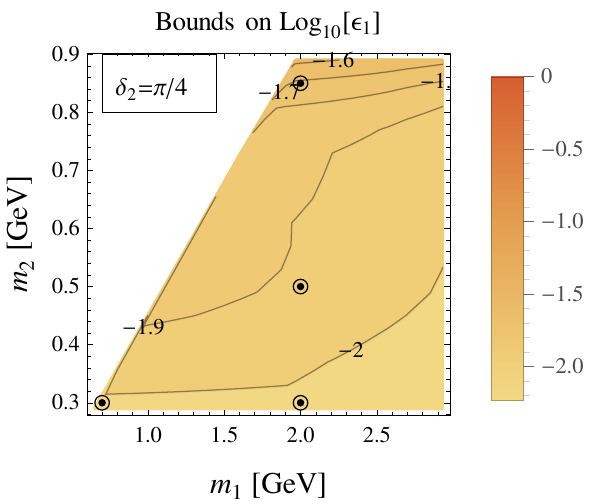}\hspace{0.1in}\includegraphics[width=0.4\textwidth]{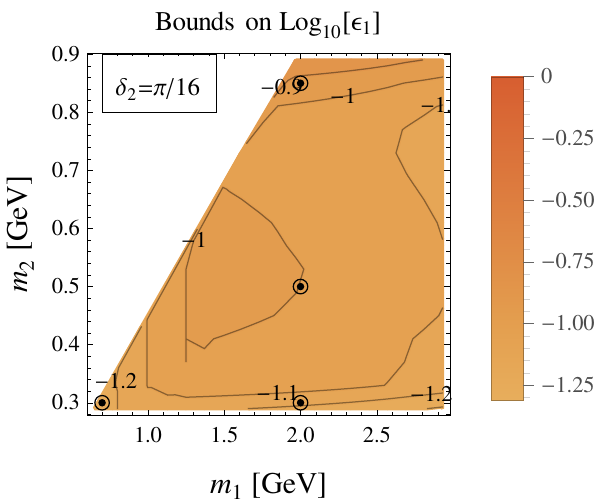}
\caption{Bounds on $\e_1$ for $\delta_2=\pi/4$ (left) and $\delta_2=\pi/16$ (right). As expected choosing lower $\delta_2$ relaxes the dominant bound from $B_s \to 4\mu$ which leads to larger $\e_1$.}%
\label{fig:e1bounds}%
\end{figure}
 
\subsection{Changes to the $B-\bar{B}$ oscillations}

Both $n_1$ and $n_2$ can cause the transition between $B_q$ and $\bar{B}_q$ for $q\in \{d,s\}$ by participating in the following diagrams: 
	\begin{center}
		\includegraphics[width=0.4\textwidth]{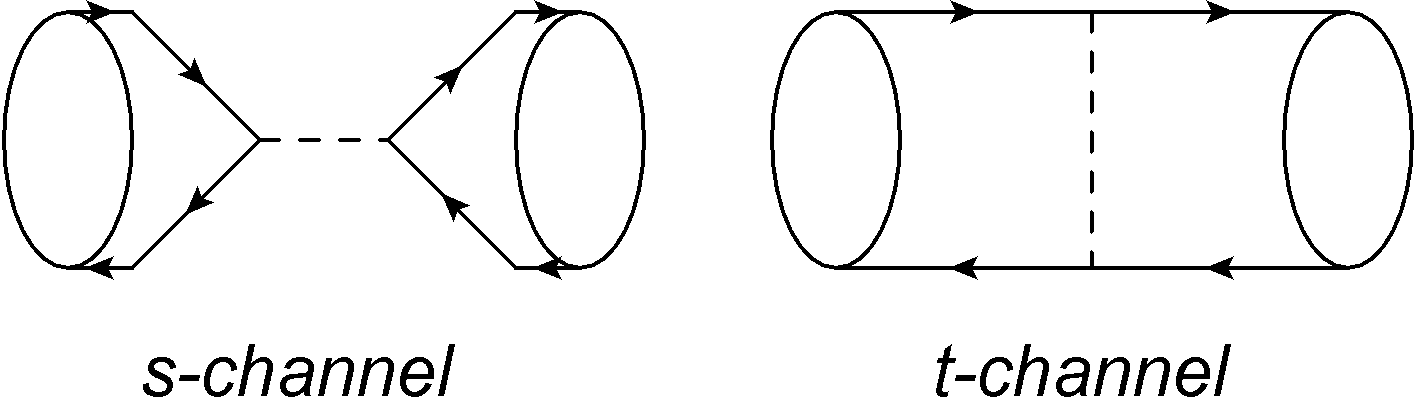}
	\end{center}
In the $s$-channel the momentum running through the $n_1$ propagator is just $m_{B_q}$. In the $t$-channel the momentum depends on the parton wave functions inside the B meson. Nevertheless, it will be on the order of $m_b-m_s$, therefore not much different from the $s$-channel and we will assume the two to be comparable. With this assumption we can extract the contribution to both $\Delta m_q = 2 M_{12} = \mathrm{Re} (\mathcal{M})/m_B$ and $\Delta \Gamma_q =  2\Gamma_{12} = \mathrm{Im} (\mathcal{M})/m_B$:
		\begin{equation}
			\begin{split}
			\delta M_{12} =& \frac{2}{3}(\mathcal{B}\lambda_p^2  m_B f_B^2)\left(\frac{\e_1^2 \cos^2(\delta_1+\alpha-\gamma)(m_{1}^2-m_{B}^2)}{(m_{1}^2-m_{B}^2)^2+ m_{1}^2 \Gamma_{1}^2(m_B^2)}+\frac{\e_2^2 \cos^2(\delta_2+\alpha-\gamma)(m_{2}^2-m_{B}^2)}{(m_{2}^2-m_{B}^2)^2+ m_{2}^2 \Gamma_{2}^2(m_B^2)}\right)\\
			\delta \Gamma_{12} =& \frac{1}{3}(\mathcal{B}\lambda_p^2 m_B f_B^2)\left(\frac{-\e_1^2 \cos^2(\delta_1+\alpha-\gamma) m_{1}\Gamma_{1}(m_B^2)}{(m_{1}^2-m_{B}^2)^2+ m_{1}^2 \Gamma_{1}^2(m_B^2)}+\frac{-\e_2^2 \cos^2(\delta_2+\alpha-\gamma)m_{2}\Gamma_{2}(m_B^2)}{(m_{2}^2-m_{B}^2)^2+ m_{2}^2 \Gamma_{2}^2(m_B^2)}\right),
			\end{split}
		\end{equation}
where $\mathcal{B} \sim \mathcal{O}(1)$ is the bag parameter associated with the scalar operator $(b(1+\gamma_5)\bar{s})^2$. We use the scenario I from \cite{Lenz:2012az,Lenz:2010gu} to evaluate the theoretical uncertainties connected with these measurements. The allowed deviations we are going to use are in table~\ref{tab:mix}. Since the actual deviations caused by this New Physics are quite small, it is unnecessary to study the relative CP violating phases $\phi_d$ and $\phi_s$. 
\begin{table}%
\centering
\begin{tabular}{c|cc}
Observable & Current experimental value & $2\sigma$ Allowed NP contribution \\\hline
$\Delta m_d$ & $3.3\times 10^{-13}\;\mathrm{GeV}$ & $(-9.7,7.0)\times 10^{-14}\;\mathrm{GeV}$ \\
$\Delta \Gamma_d$ & $2.5\times 10^{-15}\;\mathrm{GeV}$ & $(-0.7,1.6)\times 10^{-15}\;\mathrm{GeV}$ \\\hline
$\Delta m_s$ & $1.2\times 10^{-11}\;\mathrm{GeV}$ & $(-1.6,2.1)\times 10^{-12}\;\mathrm{GeV}$ \\
$\Delta \Gamma_s$ & $6.6\times 10^{-14}\;\mathrm{GeV}$ & $(-2.4,3.7)\times 10^{-14}\;\mathrm{GeV}$\\\hline
\end{tabular}
\caption{Allowed deviations from $B_q-\bar{B}_q$ mixing observables}
\label{tab:mix}
\end{table}
Nevertheless, the changes to $\Delta m_q$ and $\Delta \Gamma_q$ are at most on the order $\mathcal{O}(10^{-3})$, given other constraints on $\e_1$ and $\e_2$.

\subsection{Collider Constraints: Higgs Decays and $pp\to\bar{b}b n_i$}

Since the DS directly couples to the Higgs sector of the Standard Model, we  consider the constraints that would arise from collider studies. One of these is the invisible Higgs width, another is the associated production of $n_i$ with a $\bar{q}q$ pair: $pp \to \bar{q}q n_i$.

If the $\bar{b}b$ is produced with energy much bigger than $E_{bb} \gg m_i$, it is quite possible to radiate $n_i$. The rate of radiation of a soft $n_i$ is on the order:
\begin{equation}
\sigma(pp\to\bar{q}q + n_i) = \frac{y_q^2 \e_i^2 f(\delta_i)^2}{4\pi}  \sigma(pp\to \bar{q}q),
\end{equation}
where $f(\delta_i)$ is either $\cos(\delta_i)$ or $\sin(\delta_i)$ depending on the type of the quark. Since $\e_1 \gg \e_2$, $n_1$ makes the dominant contribution. A radiated $n_i$ would promptly decay into $2n_2, 3n_2$ or $4n_2$ and these would then appear as multiple-muons, jets or muon rich jets, depending on $m_2$ and $\delta_2$. The investigation of this phenomenon is quite beyond the scope of this paper and would be great topic of future work. Nevertheless, for $\delta_2=\pi/4$, $\e_1 \sim 10^{-2}$, which means that a pair of b quarks would radiate a soft $n_1$ with probability of about $5\times 10^{-8}$. Since hard $\bar{b}b$ pair has a cross-section of about $11\;\mathrm{nb}$, this makes the cross-section for radiative $\bar{b}bn_1$ on the order $0.5\;\mathrm{fb}$, meaning there are about forty events in the $20\;\mathrm{fb}$ dataset. Therefore we believe this process does not represent a challenge to our model, so far.

As a result of the mass mixing our model allows processes such as $h \to n_i n_j$. Since $m_h \gg m_1,m_2$, the available phase space is about the same whether we consider $h\to 2n_1, n_1 n_2$, or $2n_2$. The partial width for $h\to n_i n_j$ is:
\begin{equation}
\Gamma(h \to \mathrm{NP}) = \frac{1}{16\pi m_h}\sum_{ij} \frac{1}{S}\Lambda_{hij}^2,
\end{equation}
where $S$ stands for the necessary symmetry factor. If we impose a fairly loose constraint $\Gamma(h \to \mathrm{NP}) \lesssim \frac{1}{2}\Gamma(h \to \mathrm{SM}) \sim 2\;\mathrm{MeV}$, which would correspond to about $30\%$ branching fraction for invisible decays of the Higgs boson. Thus we obtain a bound:
\begin{equation}
\frac{1}{2}\Lambda_{h11}^2+\Lambda_{h12}^2+\frac{1}{2}\Lambda_{h22}^2 \lesssim 16\pi (125\;\mathrm{GeV})(2\;\mathrm{MeV}) \sim \left(3.5\;\mathrm{GeV}\right)^2
\end{equation} 
As we have shown in equation~\ref{eq:cubics}, the expected size of these operators is roughly $\sum \e_i \Lambda_{ijk}$. As we will see, $\e_1 \lesssim 10^{-1}$ or lower, and $\e_2\lesssim 10^{-3}$. Since the largest $\Lambda_{ijk}$ is $\Lambda_{111} \lesssim \sqrt{16 \pi} \times \;\mathrm{GeV} \sim 20\;\mathrm{GeV}$ -- this puts a slight constraint on $\e_1$ for the more massive $n_1$.

The result of applying all the bounds on $\e_1$ mentioned so far is summarized in figure~\ref{fig:e1bounds}.

\section{New Decay Channels of $\mathbf{B_q}$}
\label{sec:Bresults}
This section presents decay channels of B mesons into multi-particle final states that have not been experimentally constrained. Our model provides a way to achieve rather high branching fractions for these modes. All of the results are achieved by saturating the bounds on $\e_1$ and $\e_2$ from section~\ref{sec:constraints}.

\subsubsection{Five Particle Final States}
Instead of completely annihilating, the flavor changed constituent quarks of B meson might form another scalar or vector meson which appears in the final state. Therefore instead of $B_q \to 2n_2$ we might also observe $B_q \to M+2n_2$, where $M$ stands for any meson. In the future, we will use $S$ for a scalar or pseudo-scalar meson and $V$ for vector or pseudo-vector meson.
 
\begin{figure}[ht]%
\centering
\includegraphics[width=0.4\columnwidth]{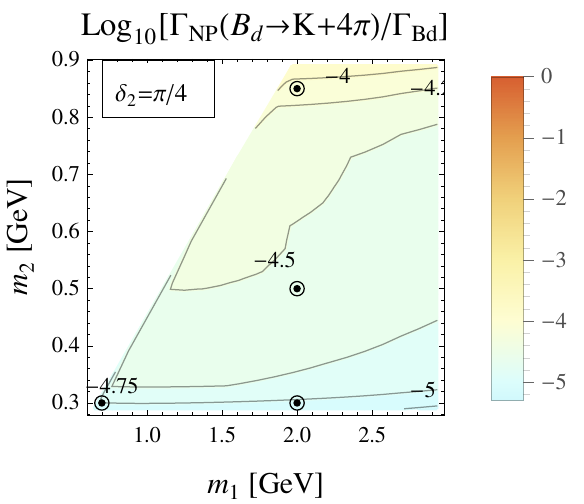}\hspace{0.05in}\includegraphics[width=0.4\columnwidth]{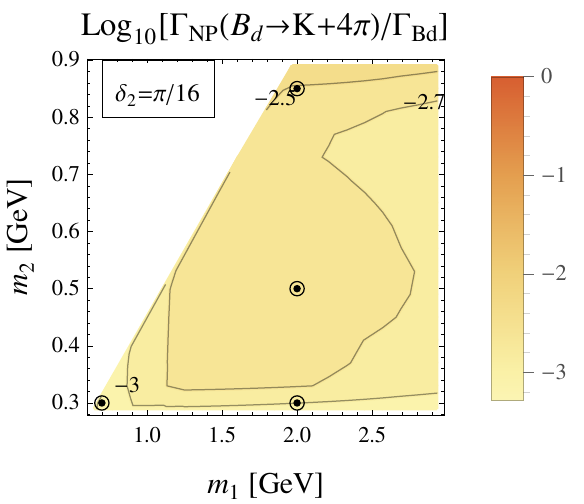}\\
\includegraphics[width=0.4\columnwidth]{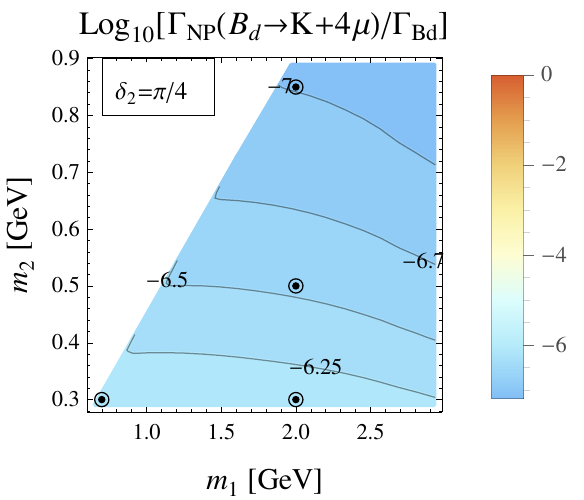}\hspace{0.05in}\includegraphics[width=0.4\columnwidth]{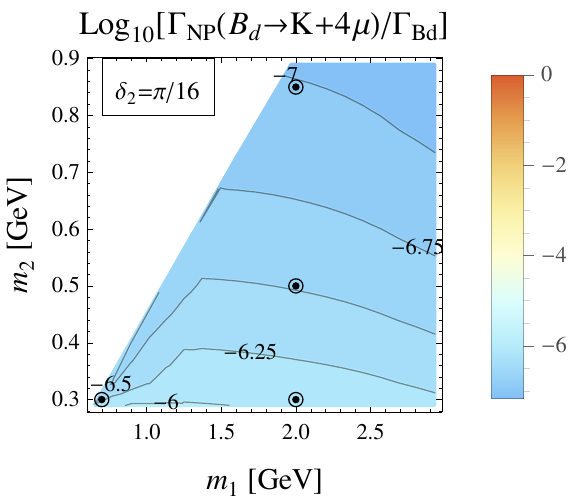}
\caption{Five Particle decays $B_d \to K+2n_2$. The purely muonic decay modes could be observable given the bounds on $B_q \to 4\mu$.}
\label{fig:5PFSS}
\end{figure}

\begin{figure}[ht]%
\centering
\includegraphics[width=0.4\columnwidth]{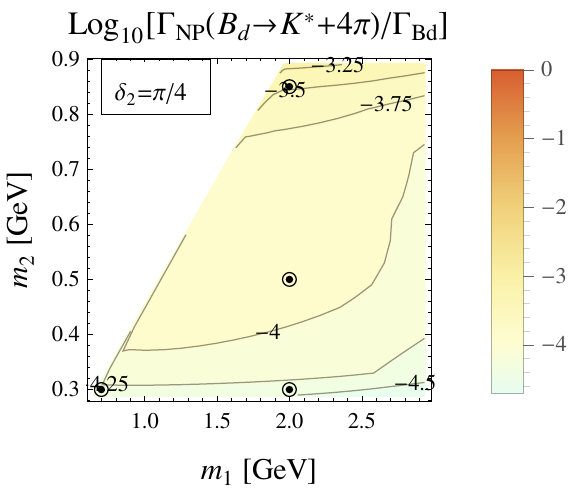}\hspace{0.05in}\includegraphics[width=0.4\columnwidth]{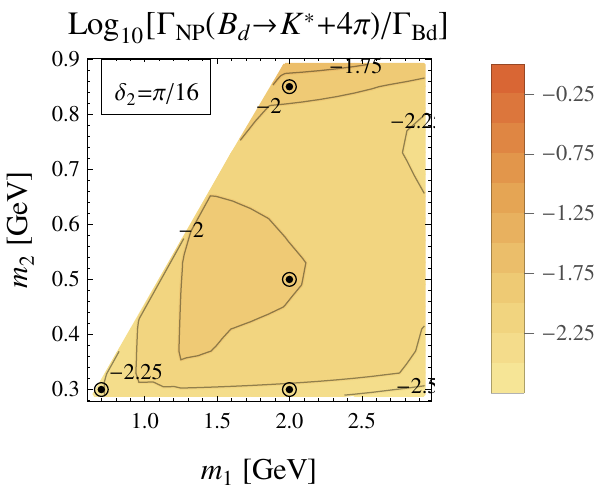}\\
\includegraphics[width=0.4\columnwidth]{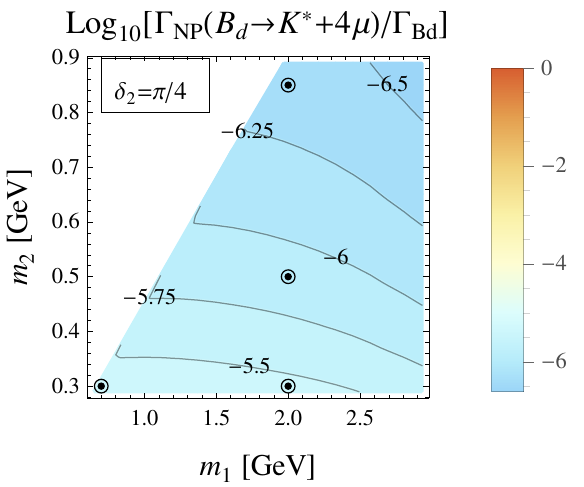}\hspace{0.05in}\includegraphics[width=0.4\columnwidth]{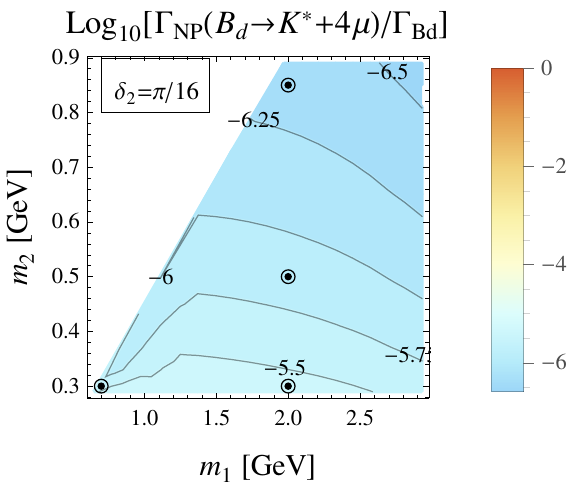}
\caption{Five Particle decays $B_d \to K^*+2n_2$. The purely muonic decay modes should be observable given the bounds on $B_q \to 4\mu$. The allowed branching fraction for $B_d \to K^{(*)}+4\pi$ are very large and should be easy to constrain with experimental measurement.}
\label{fig:5PFSV}
\end{figure}

Under current constraints on $\e_1$ these decay modes have almost absurdly large branching fractions in the mesonic decay channels of $B_d$. Just as we have seen in the comparison between the 2PFS and 3PFS, the contribution from on-shell $n_1$ or $n_2$ can be large enough that these processes effectively become decays into two particle states, $B_q \to M n_i^*$, with $n_i^*$ slightly off-shell. We use the following width to obtain our predictions for final branching fractions we plot in figures~\ref{fig:5PFSS} and~\ref{fig:5PFSV} :  
\begin{multline}
\frac{d\Gamma(B_d \to K+2n_2)}{dq^2} = \frac{1}{128 \pi^2} \frac{\lambda_{122} \e_1^2 \lambda_{s}^2 m_2^2 \cos^2(\alpha-\gamma-\delta_1)}{m_b^2 m_{B_d}^3}(m_{B_d}^2-m_K^2)^2 \left|f_0^{B\to K}(q^2)\right|^2\\\frac{\sqrt{1-4m_2^2/q^2}\sqrt{\left(m_{B_d}^2-(|q|+m_K)^2\right)\left(m_{B_d}^2-(|q|-m_K)^2\right)}}{(q^2-m_1^2)^2+m_1^2 \Gamma_1^2(q^2)}
\end{multline}
\begin{multline}
\frac{d\Gamma(B_d \to K^*+2n_2)}{dq^2} = \frac{1}{16 \pi^2} \frac{\lambda_{122} \e_1^2 \lambda_{s}^2 m_2^2 \cos^2(\alpha-\gamma-\delta_1)}{m_b^2 m_{B_d}^3} \left|A_0^{B\to K^\star}(q^2)\right|^2\\\frac{\sqrt{1-4m_2^2/q^2}\left(\left(m_{B_d}^2-(|q|+m_{K^\star})^2\right)\left(m_{B_d}^2-(|q|-m_{K^\star})^2\right)\right)^{3/2}}{(q^2-m_1^2)^2+m_1^2 \Gamma_1^2(q^2)}
\end{multline}

\begin{figure}%
\centering
\includegraphics[width=0.4\columnwidth]{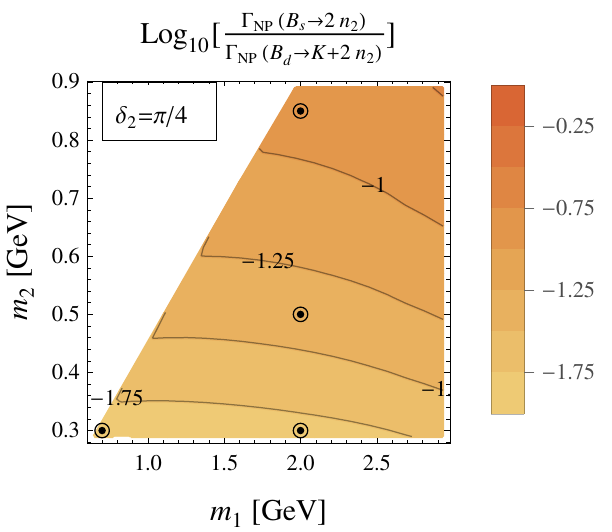}\hspace{0.1in}\includegraphics[width=0.4\columnwidth]{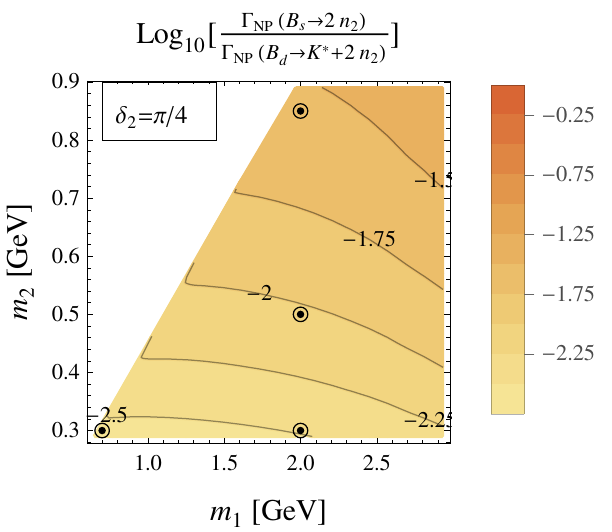}
\caption{Comparison between four and five particle final states. This ratio is independent of $\e_1$, $\e_2$, $\delta_1$, $\delta_2$ and $\lambda_{122}$. You can see it does depend on $m_1$ and $m_2$. We can see that the five particle final states are preferred.}%
\label{fig:45PFS}%
\end{figure}

These modes present a great way to identify this particular model of the DS. Since all decays proceed through a penguin operator, the annihilation decays of $B_d$ are suppressed by a factor $|V_td/V_{ts}|^2 \sim 0.04$. Nevertheless, the decays $B_d \to K +\;(\mathrm{Dark\; Sector})$ proceed through the $b\to s$ penguin operator and therefore are not suppressed. Figure~\ref{fig:45PFS} shows the ratio $\Gamma_{NP}(B_s \to 2n_2)/\Gamma_{NP}(B_d \to K+ 2n_2)$, which is independent of all the couplings in the DS: $\e_1$, $\e_2$, $\delta_1$, $\delta_2$ and $\lambda_{122}$. This ratio should then only be dependent on $m_1$, $m_2$ and the kinematic of the Standard Model bound states and forms an independent check on the model in decay modes of two \emph{different} particles. 

At first, it may be surprising that $B_d \to K+4\pi$ has a larger rate compared to $B_s \to 4\pi$. However, since $m_B - m_K \sim m_B$ the only phase-space suppression comes from the $(4\pi)^{-1}$ factor. However, adding the Kaon allows $n_1$ to contribute on-shell and the form factor for $B_d \to K$ is typically larger than the annihilation form factor:
\begin{align}
\langle K | b\bar{s}|B\rangle &\sim (m_B^2-m_K^2)^2 \mathcal{O}(1) \\
\langle 0 | b\bar{s}|B\rangle &\sim f_B m_B^2
\end{align} 
Since the size of the phase-space is of the order $m_B$, ratio of these two is roughly $f_B/m_B \sim 20$. We will see that this trend persists and six and seven particle final states will also have comparable rates. 

\subsection{Six, Seven and Eight Particle Final States}

The decay channel $B_s \to n_1^*$ does not have to proceed to $2n_2$ it can also turn into $n_1 n_2 \to 3n_2$ respectively $2n_1\to 4n_2$ and so on. The later two options produce six and seven or eight and nine particle states, respectively. The even number particle states come from annihilation diagrams. Since we have designed the DS to sit close to the strongly coupled regime, additional branchings do not cost much and we expect these processes to have comparable branching fractions. Equations~\ref{eq:6PFS}, \ref{eq:7PFSS}, \ref{eq:7PFSV} and \ref{eq:8PFS} show the widths for these processes.
\begin{multline}
\frac{d\Gamma(B_s \to 3n_2)}{dq^2}=\frac{1}{6\pi}\frac{\lambda_{112}\lambda_{122} \e_1^2 \lambda_{s}^2 \cos^2(\alpha-\gamma-\delta_1) m_2^4 m_{B_s} f_{B_s}^2 }{m_b^2}\times\\ \times\frac{\sqrt{1-4m_2^2/q^2}}{(m_{B_s}^2-m_1^2)^2 + m_1^2 \Gamma_1(m_{B_s}^2)}
\frac{\sqrt{\left(m_{B_s}^2-(|q|+m_2)^2\right)\left(m_{B_s}^2-(|q|-m_2)^2\right)}}{(q^2-m_1^2)^2 + m_1^2 \Gamma_1(q^2)}
\label{eq:6PFS}
\end{multline}

\begin{multline}
\frac{d^2\Gamma(B_d \to K + 3n_2)}{dq^2_{123} dq^2_{12}}= \frac{1}{12\pi^3 } \frac{\lambda_{112}\lambda_{122} \e_1^2 \lambda_{s}^2 \cos^2(\alpha-\gamma-\delta_1) m_2^4 m_{B_d}}{m_b^2 q^2_{123}}\left(1-\frac{m_K^2}{m_{B_d}^2}\right)^2\left|f_0^{B\to K}(q_{123}^2)\right|^2\times\\
\times\sqrt{1-4m_2^2/q_{12}^2}\frac{\sqrt{\left(m_{B_d}^2-(m_K+|q_{123}|)^2\right)\left(m_{B_d}^2-(m_K-|q_{123}|)^2\right)}}{(m_1^2-q_{123}^2)^2+m_1^2\Gamma_1^2(q_{123}^2)}\times\\\times\frac{\sqrt{\left(q^2_{123}-(m_2+|q_{12}|)^2\right)\left(q^2_{123}-(m_2-|q_{12}|)^2\right)}}{(m_1^2-q_{12}^2)^2+m_1^2\Gamma_1^2(q_{12}^2)}
\label{eq:7PFSS}
\end{multline}

\begin{multline}
\frac{d^2\Gamma(B_d \to V + 3n_2)}{dq^2_{123} dq^2_{12}}=\frac{1}{3\pi^3 } \frac{\lambda_{112}\lambda_{122} \e_1^2 \lambda_{s}^2 \cos^2(\alpha-\gamma-\delta_1) m_2^4 }{m_b^2 q^2_{123} m_{B_d}^3}\left|A_0^{B\to K^*}(q_{123}^2)\right|^2\times\\
\times\sqrt{1-4m_2^2/q_{12}^2}\frac{\left(\left(m_{B_d}^2-(m_K^{*}+|q_{123}|)^2\right)\left(m_{B_d}^2-(m_{K^*}-|q_{123}|)^2\right)\right)^{3/2}}{(m_1^2-q_{123}^2)^2+m_1^2\Gamma_1^2(q_{123}^2)}\times\\\times\frac{\sqrt{\left(q^2_{123}-(m_2+|q_{12}|)^2\right)\left(q^2_{123}-(m_2-|q_{12}|)^2\right)}}{(m_1^2-q_{12}^2)^2+m_1^2\Gamma_1^2(q_{12}^2)}
\label{eq:7PFSV}
\end{multline}

\begin{multline}
\frac{d^2\Gamma(B_s \to 4n_2)}{dq_{12}^2 dq_{34}^2} = \frac{1}{24\pi^2 m_b^2} \frac{\lambda_{111}\lambda_{122}^2 \e_1^2 \lambda_{s}^2 \cos^2(\alpha-\gamma-\delta_1) m_1^2 m_2^4 m_{B_s} f_{B_s}^2 }{(m_1^2-m_{B_s}^2)^2 + m_1^2 \Gamma_1^2(m_{B_s}^2)}\times\\\times\sqrt{\left(1-4m_2^2/q_{12}^2\right)\left(1-4m_2^2/q_{34}^2\right)}\;\times\\
\times\frac{\sqrt{\left(m_{B_s}^2-(|q_{12}|+|q_{34}|)^2\right)\left(m_{B_s}^2-(|q_{12}|-|q_{34}|)^2\right)}}
{\left((m_1^2-q_{12}^2)^2+m_1^2 \Gamma_1^2(q_{12}^2)\right)\left((m_1^2-q_{34}^2)^2+m_1^2 \Gamma_1^2(q_{34}^2)\right)}
\label{eq:8PFS}
\end{multline}

\begin{figure}[ht]%
\centering
\includegraphics[width=0.35\columnwidth]{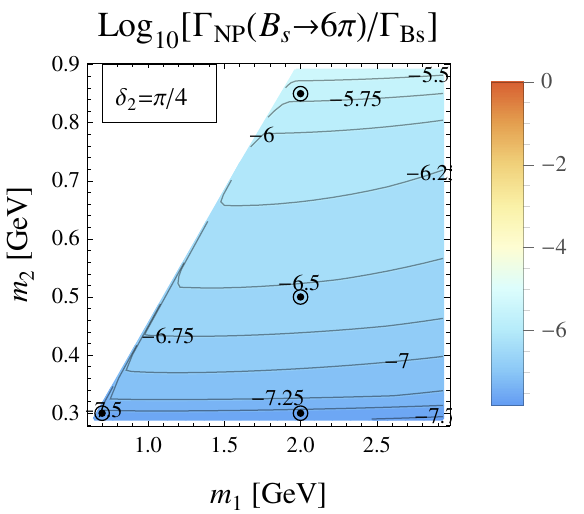}\hspace{0.05in}\includegraphics[width=0.35\columnwidth]{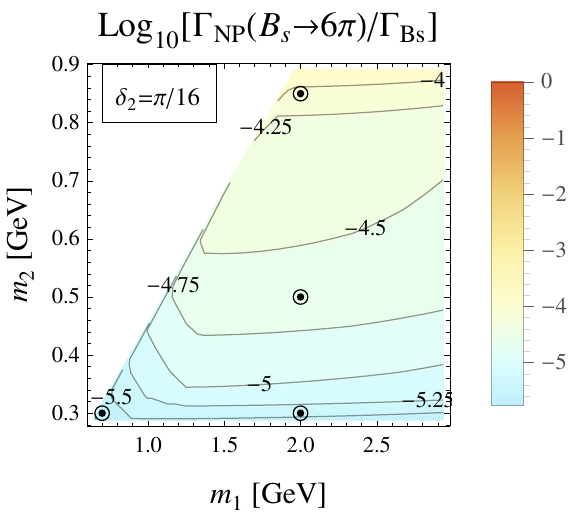}\\
\includegraphics[width=0.35\columnwidth]{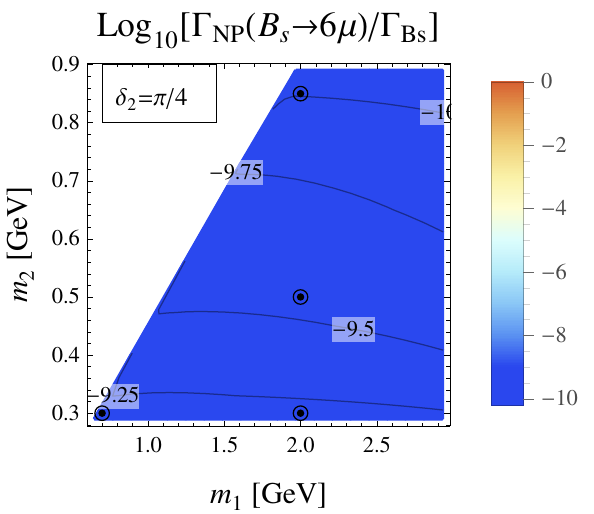}\hspace{0.05in}\includegraphics[width=0.35\columnwidth]{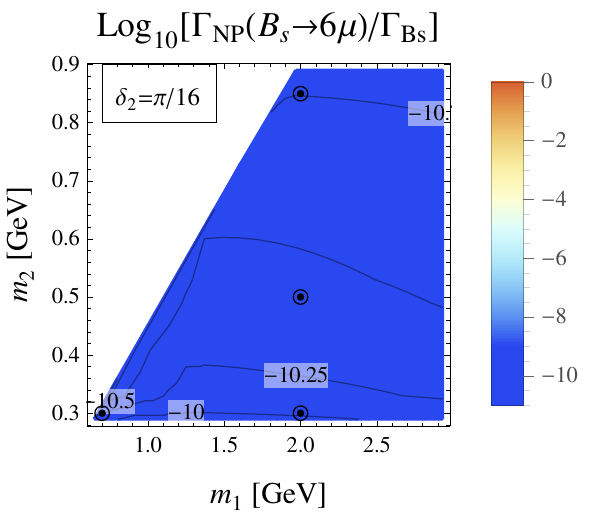}\\
\includegraphics[width=0.35\columnwidth]{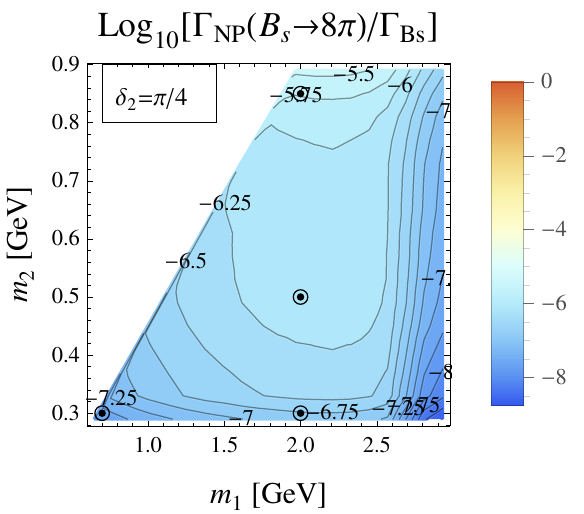}\hspace{0.05in}\includegraphics[width=0.35\columnwidth]{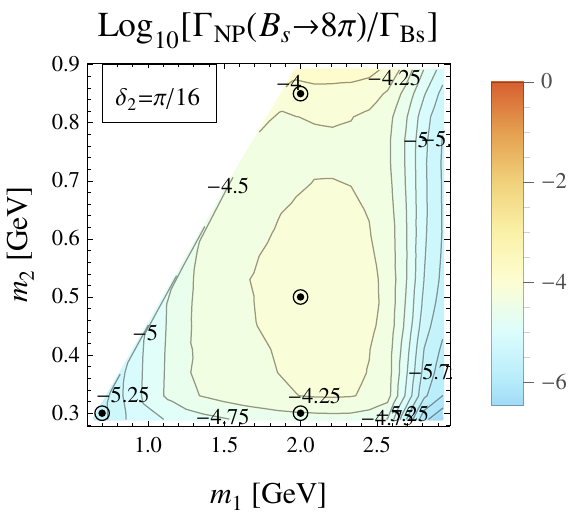}\\
\includegraphics[width=0.35\columnwidth]{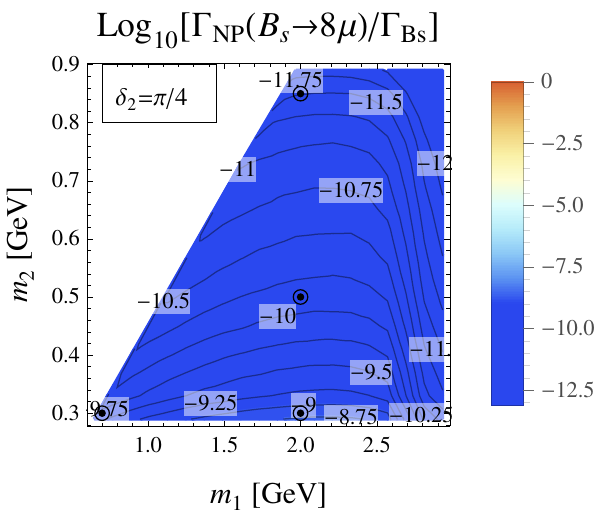}\hspace{0.05in}\includegraphics[width=0.35\columnwidth]{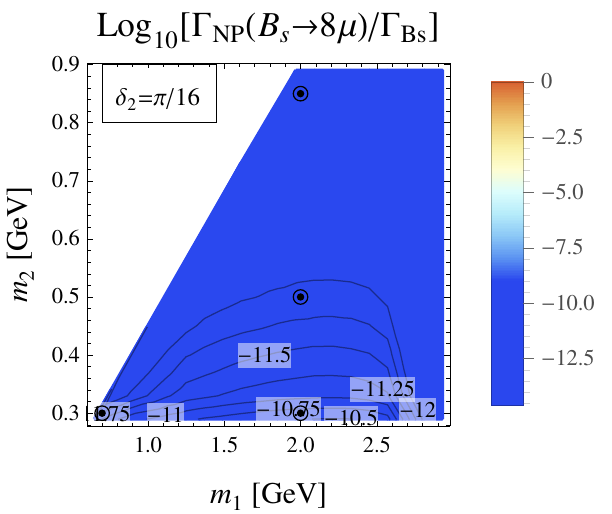}
\caption{Six and Eight Particle Final States. The couplings between $n_1$ and $n_2$ are strong and we expect that additional particles in the final state do not significantly change the width for the process. It is clear that searching in the pure muonic channel would be fruitless. However, decays into purely hadronic decay channels should be abundant. }
\label{fig:6and8PFS}
\end{figure}

\begin{figure}[ht]%
\centering
\includegraphics[width=0.35\columnwidth]{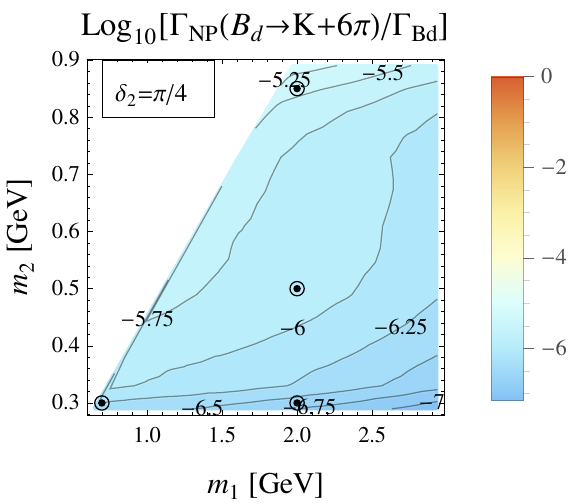}\hspace{0.05in}\includegraphics[width=0.35\columnwidth]{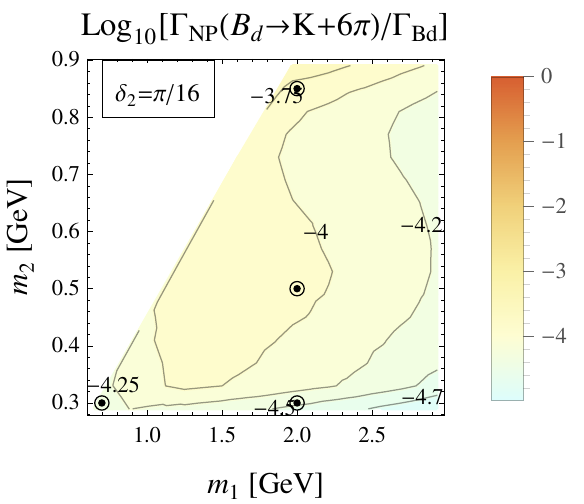}\\
\includegraphics[width=0.35\columnwidth]{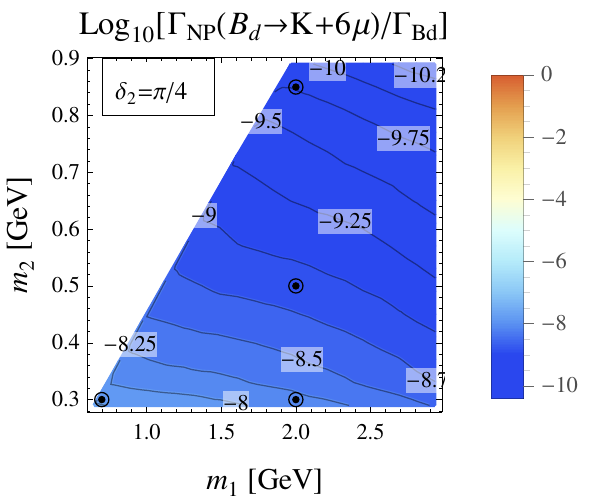}\hspace{0.05in}\includegraphics[width=0.35\columnwidth]{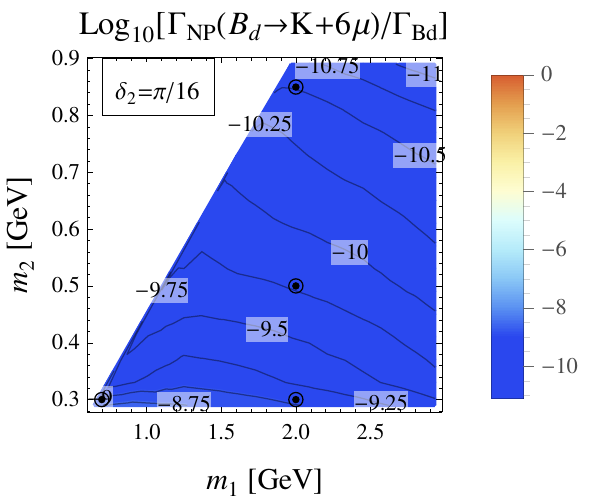}\\
\includegraphics[width=0.35\columnwidth]{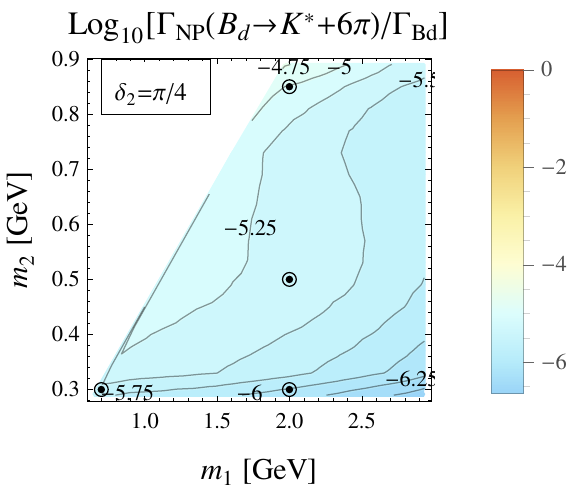}\hspace{0.05in}\includegraphics[width=0.35\columnwidth]{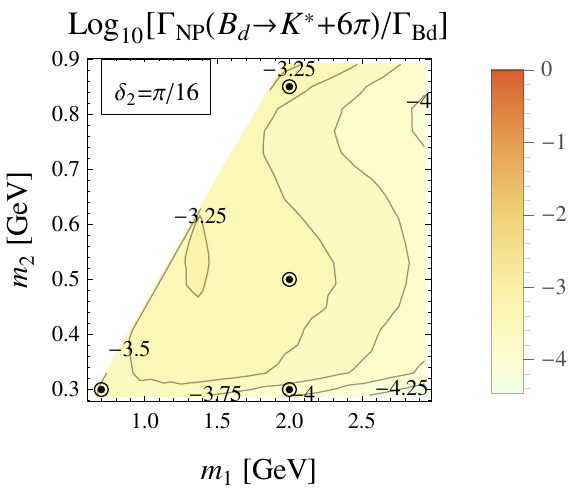}\\
\includegraphics[width=0.35\columnwidth]{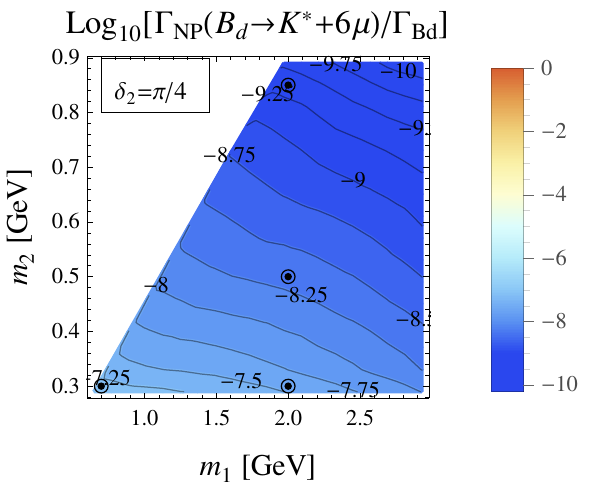}\hspace{0.05in}\includegraphics[width=0.35\columnwidth]{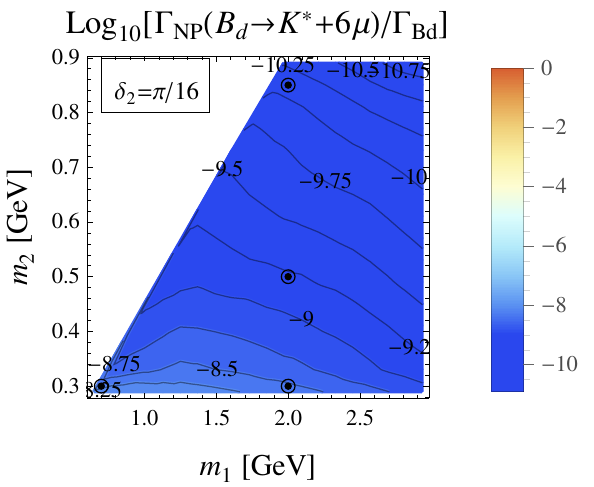}
\caption{Seven Particle Final States. Additional meson in the final state can increase the branching fraction for the same \HS decay. }
\label{fig:7PFS}
\end{figure}

The purely muonic final states are highly suppressed by the small muon branching fraction $BF(n_2 \to \mu^+\mu^-)$. Branching fractions on the order $\mathcal{O}(10^{-10})$ and lower rule out the possibility that discovery of New Physics will ever happen in the purely muonic final states. Instead we should turn our attention to hadronic decays as is apparent from figures~\ref{fig:6and8PFS} and~\ref{fig:7PFS}.

\section{Conclusion}
\label{sec:conclusions}
We have considered a very simple model of the dark sector. By coupling this model to the Standard Model through a two Higgs doublet generalization of the Higgs portal we allow charmless high particle multiplicity decay modes of B mesons. The B mesons decays include new exotic scalars, which tend to decay into into pairs of pions much more often than into pairs of muons.  Thus, existing searches involving muons in the final state still allow a large parameter space for significant branching fractions into final states with multiple pions. 
Although hadronic decays of B mesons are typically harder to constrain and the Standard Model backgrounds are hard to predict compared to their leptonic counterparts, our model offers branching fractions so large ($\sim 10^{-3}$) an experimental study should be able to significantly constrain the parameter space of our model. The signature of this model is a correlation between these exotic decay modes for $B_d$, $B_u$ and $B_s$ as well as presence of pion resonances that are only seen in high multiplicity  $B$-hadron decays.

There are several directions in which our study could be expanded. We have not covered all the decay modes this model allows. We estimate that the branching fractions for higher and higher multiplicity final states begin to drop when the phase space available to the final state particles becomes small. In particular the final number of $n_2$s in the final state cannot exceed $m_B/m_2 \lesssim 17$. Investigating these spectacular $B \to \approx30\mu$ decay modes might be fun.

Since our DS is strongly coupled we believe that the effect of quartic and cubic couplings is comparable. However, a more detailed study of this claim could prove worthwhile.

Full collider phenomenology of this model is also beyond the scope of this paper and would benefit from future attention. Some of the consequences of this model have already been described in terms of muon-jets and photon-jets.

Finally, in order to maintain some predictive power for the branching ratio $\Gamma(n_2 \to \mu\mu)/\Gamma(n_2 \to \pi\pi)$, we have maintained $m_2 < 2m_K$. However, there is no physical reason this is the case. Once $m_2 > 2m_K$ not only it is harder to make any accurate predictions but also additional decay modes such as $n_2 \to KK$ become important signatures to look for. 

\section{Acknowledgments}
We would like to thank Tuhin Roy for very helpful discussions. Our work was supported, in part, by the US Department of Energy under contract numbers DE-FGO2-96ER40956. We thank the participants in the workshop   ``New Physics from Heavy Quarks in Hadron Colliders"  for helping to inspire this project. This workshop   was sponsored by the University of Washington and supported by the DOE. JS would also like to acknowledge partial support from a DOE High Energy Physics Graduate Theory Fellowship.

\appendix
\section{Wide $n_1$}
\label{sec:widen1}
With the definition $\Lambda_{122} = \sqrt{16\pi \lambda_{122}} m_2$, let us have a look at the loop correction to the $n_1 n_2^2$ vertex:
\begin{equation}
	\vcenter{\hbox{\includegraphics[width=0.2\textwidth]{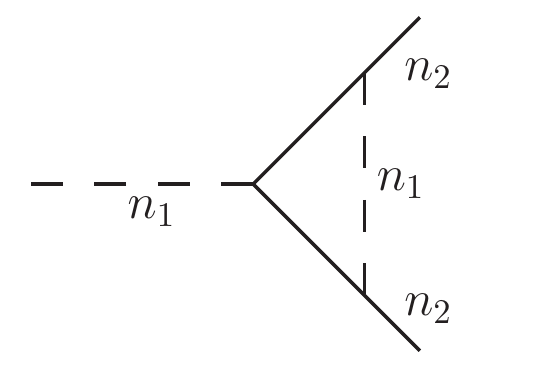}}} = \Lambda_{122}^3 \int \frac{dq^2}{16\pi^2}\frac{q^2}{(q^2-m_2^2)^2(q^2-m_1^2)} = \Lambda_{122}\frac{\lambda_{122}}{2\pi} f\left(\frac{m_2}{m_1}\right)  
\end{equation}
		The function $f(m_2/m_1)$, ranges between $0$ and $1$ and so taking $\lambda_{122} \sim 1$ already seems to ensure the one-loop correction is subleading to the tree-level amplitude.  
\begin{center}
	\includegraphics[width=0.4\textwidth]{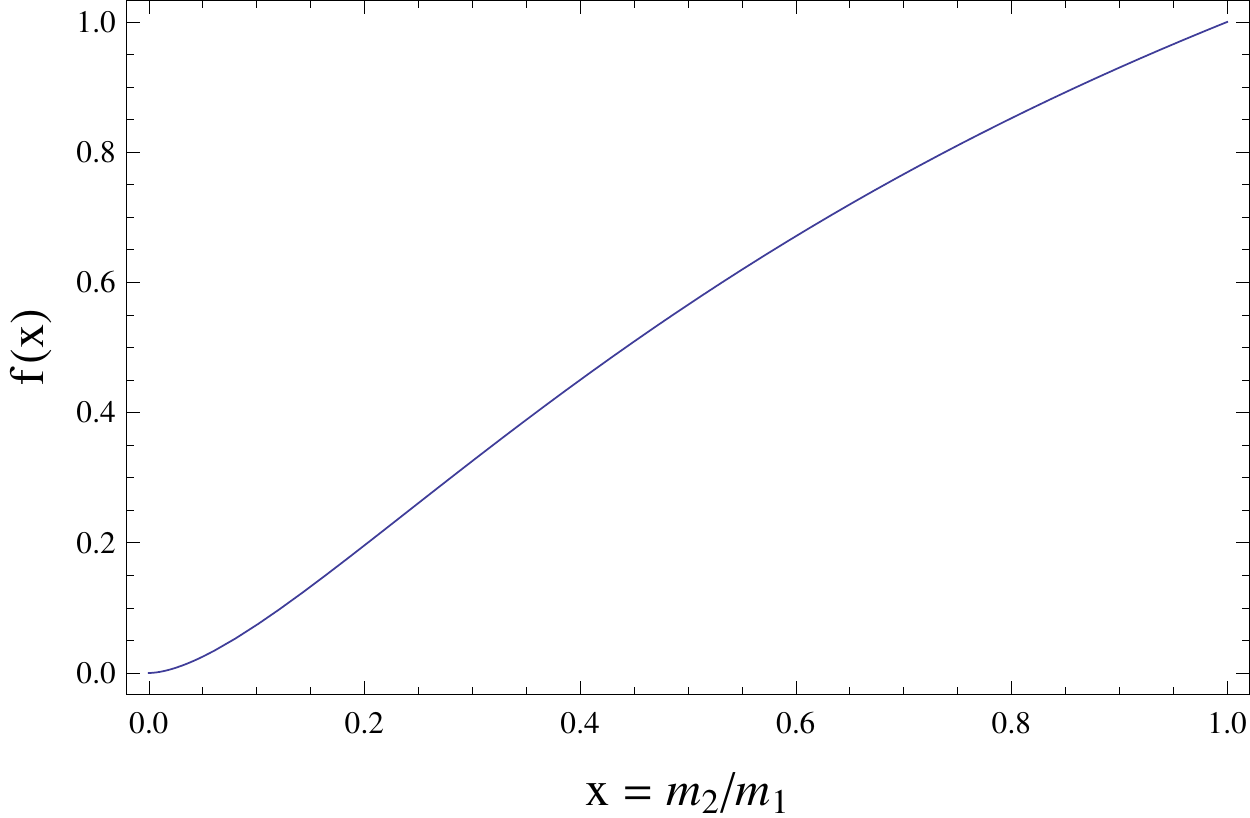}
\end{center}

However, the estimate for one-loop correction to the mass of $n_2$ becomes of the order of $m_2$ for a cut-off scale $\Lambda$:
\begin{equation}
\frac{\Lambda_{122}^2}{16 \pi^2}\ln\left(\frac{\Lambda^2}{m_2^2}\right) = \frac{\lambda_{122} m_2^2}{\pi}\ln\left(\frac{\Lambda^2}{m_2^2}\right) = m_2^2,
\end{equation}
which implies:
\begin{equation}
\Lambda = m_2 \mathrm{exp}\left(\frac{\pi}{2\lambda_{122}}\right)
\end{equation}
When $\lambda_{122} = 1/3$ the cut-off scale is roughly $\Lambda \sim 100 m_2$, already quite low. Nevertheless, for the masses of $n_2$ we will consider this cut-off scale is still higher than the mass of the B mesons.
\begin{center}
	\includegraphics[width=0.4\textwidth]{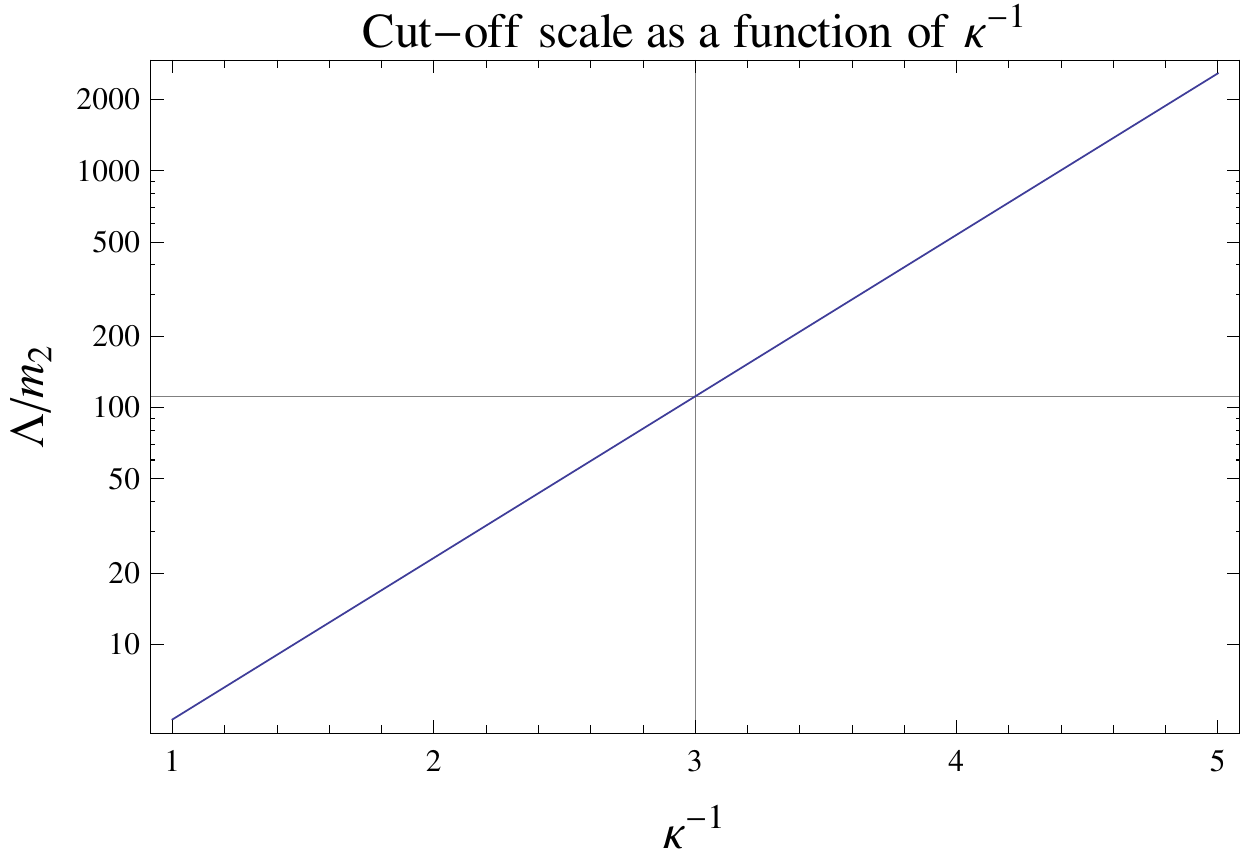}
\end{center}
However, should we be satisfied with an order percent fine-tunning, $\pi/(2\lambda_{122})$ is replaced with $100\pi/(2\lambda{122})$. This pushes available range of $\lambda_{122}$ to $\sim 30$ far out of the perturbative regime.

\section{Estimates of Branching Ratios for a Light Higgs with $\chi$PT for $m_2 < 2m_K$}
\label{sec:branch}
\begin{figure}%
\begin{center}
\includegraphics[width=0.35\textwidth]{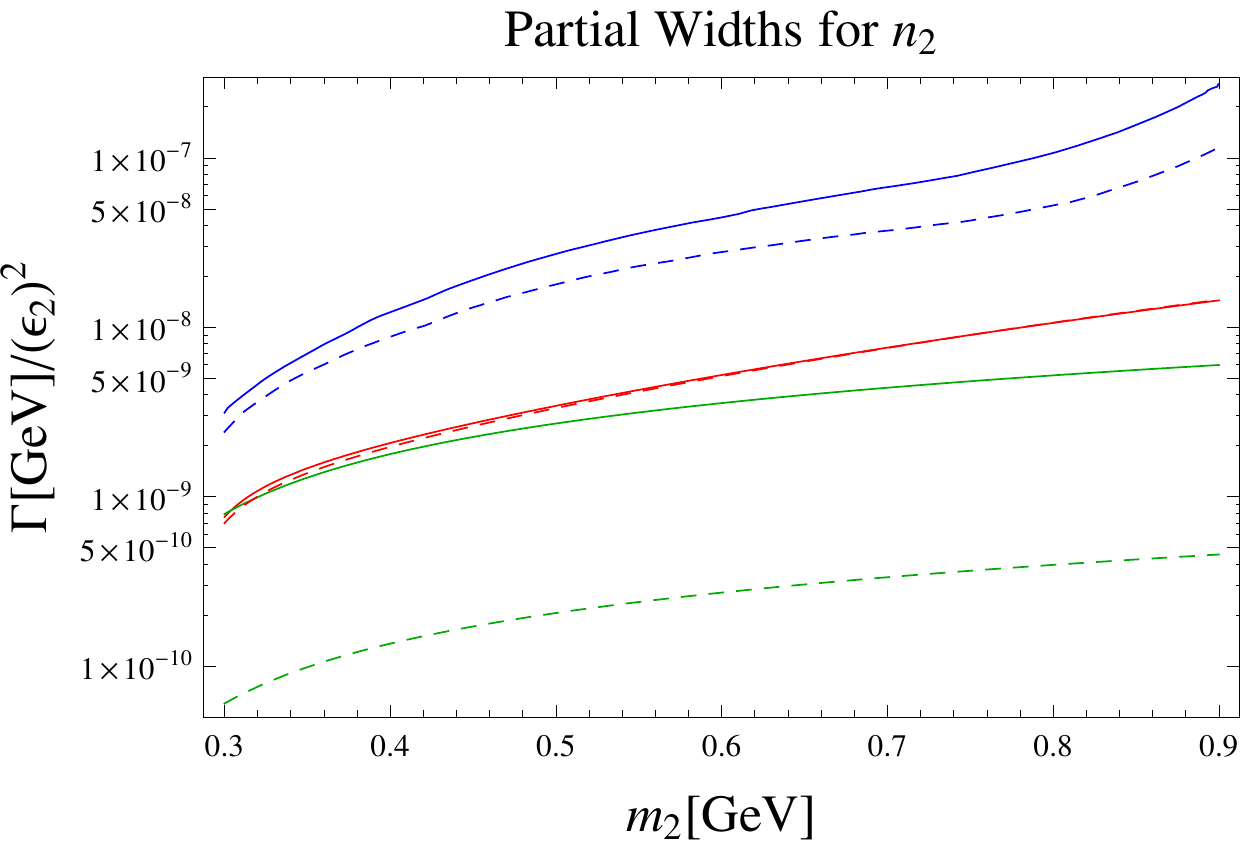}%
\caption{This plot shows partial width $\Gamma(n_2\to\mu\mu)$ in green, $\Gamma(n_2 \to \pi\pi)$ according to tree-level $\chi$PT in red and $\Gamma(n_2 \to \pi\pi)$ based on improved $\chi$PT \cite{Donoghue} in blue. The solid lines stand for $\delta_2 = \pi/4$, whereas the dashed lines mark the results for $\delta_2 = \pi/16$. The lack of change for tree-level $\chi$PT between $\delta_2 = \pi/4$ and $\delta_2 = \pi/16$ is caused by a numerical coincidence.}%
\end{center}
\label{fig:compDChi}
\end{figure}

In this mass regime we can use $\chi$PT to compare the decay widths $\Gamma(n_2 \to \ell^+\ell^-)$ and $\Gamma(n_2 \to \pi^a\pi_a)$. Although $n_2 \to \gamma\gamma$ is allowed, it is unimportant unless $m_2 < 2m_e$. Coupling of light Higgs boson is well described by \cite{Gunion:1989we} and we follow their reasoning. The basic trick is to express the effective Higgs coupling in terms of operators that are easily evaluated within $\chi$PT. The effective theory for SM Higgs coupling to gluons and quarks can be obtained from integrating out the $N_h$ heavy quark loops:
\begin{equation}
\mathcal{L}_{eff} = \frac{h}{v}\left(\frac{\alpha_s N_h}{12 \pi } G^{\mu\nu}G_{\mu\nu} - m_u\bar{u}u - m_d\bar{d}d - m_s\bar{s}s\right)  
\end{equation}
For a 2HDM this can be easily translated in the $h_u$, $h_d$ basis:
\begin{equation}
\mathcal{L}_{eff} = \frac{h_u}{v_u}\left(\frac{\alpha_s N_H^u}{12 \pi } G^{\mu\nu}G_{\mu\nu} - m_u\bar{u}u \right)+\frac{h_d}{v_d}\left(\frac{\alpha_s N_H^d}{12 \pi } G^{\mu\nu}G_{\mu\nu} - m_d\bar{d}d - m_s \bar{s}s\right)  
\end{equation}
In our case $N_h^u = 2$ and $N_h^d = 1$. As a result the $n_2$ coupling is given by:
\begin{equation}
\mathcal{L}_{eff} = \e_2 \frac{n_2}{v}\left[ \frac{ \cos \delta_2}{\sin \beta}\left(\frac{2\alpha_s}{12 \pi } G^{\mu\nu}G_{\mu\nu} - m_u\bar{u}u \right)+\frac{\sin \delta_2 }{\cos\beta}\left(\frac{\alpha_s}{12 \pi } G^{\mu\nu}G_{\mu\nu} - m_d\bar{d}d - m_s \bar{s}s\right)  \right]
\end{equation}
This is very similar to the trace of the stress-energy tensor for the gluons and fermions of this effective theory:
\begin{equation}
\theta^\mu_{\;\mu} = -\frac{9\alpha_s}{8\pi}G^{\mu\nu}G_{\mu\nu}+\sum m_q \bar{q}q \rightarrow G^{\mu\nu}G_{\mu\nu} = \frac{8\pi}{9\alpha_s}\left(\sum m_q \bar{q}q - \theta^\mu_{\;\mu}\right)
\end{equation}
And so we can express the effective coupling in terms of the stress-energy tensor and quark mass operator:
\begin{equation}
\begin{split}
\mathcal{L}_{eff} =  -\e_2 \frac{n_2}{v}\left[ \frac{2 N_E}{27} \theta^\mu_{\;\mu} + \left(\frac{ \cos \delta_2}{\sin \beta} - \frac{2 N_E}{27} \right)m_u\bar{u}u+ \left(\frac{ \sin\delta_2}{\cos \beta} - \frac{2 N_E }{27} \right)\left(m_d\bar{d}d+m_s\bar{s}s\right)\right]
\end{split}
\end{equation} 
where the effective number of heavy flavors $N_E$ depends on the couplings:
\begin{equation}
N_E = \left(2\frac{\cos\delta_2}{\sin\beta} + \frac{\sin \delta_2 }{\cos\beta}\right)
\end{equation}
On the $\chi$PT side, working with $\mathcal{L}_2$ to the leading order, the stress-energy tensor is simple:
\begin{equation}
\theta^\mu_{\;\mu} = g^{\mu\nu}\theta_{\mu\nu} = g^{\mu\nu}\frac{2}{\sqrt{-g}}\frac{\delta (\sqrt{-g}\mathcal{L}_2)}{\delta g^{\mu\nu}} = - 2 \mathcal{L}_2
\end{equation}
and so the matrix elements for transition to two pions is easy to evaluate:
\begin{equation}
\langle \pi_a\pi_b | \theta^\mu_{\;\mu}(q^2) | 0 \rangle = (q^2 + 2m_\pi^2 )\delta_{ab}
\end{equation}
We can similarly evaluate the matrix elements for the quark mass operators (since $\chi$PT predicts how pion mass depends on the quark masses):
\begin{equation}
\langle \pi_a \pi_b | \bar{q}q | 0 \rangle \rangle \pi_a \pi_b | \frac{f_\pi^2 m_0}{2}\frac{\partial \mathrm{Tr}(M\Sigma+M^\dagger\Sigma^\dagger)}{\partial m_q}|0\rangle
\end{equation}
and so ignoring electromagnetic corrections we can evaluate the necessary matrix elements:
\begin{equation}
\begin{split}
\langle \pi^+\pi^- | m_u \bar{u}u |0\rangle = m_0m_u = \frac{1}{2}\left(m_\pi^2+m_{K^+}^2-m_{K^0}^2\right)\\
\langle \pi^+\pi^- | m_d \bar{d}d |0\rangle = m_0m_d = \frac{1}{2}\left(m_\pi^2-m_{K^+}^2+m_{K^0}^2\right)
\end{split}
\end{equation}
We put all these results together to obtain the desired matrix element for $n_2 \to \pi\pi$ decay:
\begin{multline}
\langle \pi\pi | \mathcal{L}_{eff} | n_2 \rangle = -\frac{\e_2}{v}\times\\\times\left(\frac{\sin \delta_2}{\cos\beta}\right) \left(\frac{2}{27}\left(2T_{\beta\delta_2}^{-1} + 1\right)(m_2^2+m_\pi^2) + \frac{1}{2}\left(T_{\beta\delta_2}^{-1} + 1\right)m_\pi^2 + \frac{1}{2}\left(T_{\beta\delta_2}^{-1} - 1\right)\left(m_{K^+}^2-m_{K^0}^2\right) \right)
\end{multline}
This allows us to compare the relative width for hadronic and leptonic decays for $m_2 < 2m_K$. Since the muon branching fraction is proportional to $\e_2 \sin \delta_2 / \cos\beta $, the relative branching fraction is only sensitive to the two parameters: $m_2$ and the product $T_{\beta\delta_2} = \tan\beta \tan\delta_2$. We plot the comparison between the results based on \cite{Donoghue} and those obtained from using tree-level unimproved $\chi$PT in figure~\ref{fig:compDChi}.

\bibliographystyle{utphys}
\bibliography{HFDS_bib}

\end{document}